\documentclass[final,leqno,onefignum,onetabnum]{siamltex1213}

\usepackage{graphicx}
\usepackage[nolist]{acronym}
\usepackage{amsmath}
\usepackage{booktabs}
\usepackage{array}
\usepackage{diagbox}
\usepackage{tabularx}
\usepackage[nocompress]{cite}
\newcommand{\ie}{\emph{i.e.},}
\newcommand{\eg}{\emph{e.g.},}

\newcommand{\etal}{\emph{et al.}}
\newcommand{\Sec}{\S}
\newcommand{\Fig}{Figure}
\newcommand{\Table}{Table}
\newcommand{\mycaption}[2]{\caption[#1]{#1 #2}}
\newcommand{\nnz}{\mathit{nnz}}
\newcommand{\LIV}{\mathit{LIV}}
\newcommand{\nzc}{\mathit{nzc}}
\newcommand{\nzr}{\mathit{nzr}}

\newcolumntype{C}[1]{>{\centering\let\newline\\\arraybackslash\hspace{0pt}}m{#1}}
\newcolumntype{Y}{>{\centering\arraybackslash}X}

\begin{acronym}
\acro{NNZ}{number of non-zeros}
\acro{MDT}{multiplication datatype}
\acro{MV}{matrix-vector}
\acro{DSL}{domain-specific language}
\acro{MTT}{MATLAB Tensor Toolbox}
\acro{LCO}{linearised coordinate}
\acro{CO}{coordinate}
\acro{CSC}{compressed sparse-column}
\acro{DCSR}{doubly compressed-sparse row}
\acro{DCSC}{doubly compressed-sparse column}
\acro{CSR}{compressed sparse-row}
\acro{CSCNA}{compressed sparse-column no-accumulator}
\acro{CSRNA}{compressed sparse-row no-accumulator}
\acro{LIV}{linearised index value}
\acro{CSL}{C++ Standard Library}
\acro{MSD}{most-significant digit}
\acro{LSD}{least-significant digit}
\acro{ALS}{alternating least-squares}
\acro{RP}{radix permutation}
\acro{FD}{finite-difference}
\acro{NT}{numeric tensor}
\acro{BEP}{base edge probability}
\acro{SOP}{simple outer-product}
\newacroplural{LI}{linearised indices}
\newacroplural{BEP}{base edge probabilities}
\end{acronym}

\title{High Performance Rearrangement and Multiplication Routines for Sparse Tensor Arithmetic\thanks{This work was conducted at and financially supported by the University of Alberta. Support also provided by the Natural Sciences and Engineering Research Council of Canada and the Killam Trusts.}}
\author{A.~P. Harrison\footnotemark[2]\ \and D. Joseph\footnotemark[2]}

\begin{document}
\maketitle
\slugger{sisc}{xxxx}{xx}{x}{x--x}

\renewcommand{\thefootnote}{\fnsymbol{footnote}}

\footnotetext[2]{Department of Electrical and Computer Engineering,
University of Alberta, Edmonton, Alberta, Canada. (\email{adam.p.harrison@gmail.com}). Questions, comments, or corrections
to this document may be directed to that email address.}

\renewcommand{\thefootnote}{\arabic{footnote}}

\begin{abstract}
Researchers from diverse disciplines are increasingly incorporating numeric high-order data, \ie{} numeric tensors, within their practice. Just like the \ac{MV} paradigm, the development of multi-purpose, but high-performance, sparse data structures and algorithms for arithmetic calculations, \eg{} those found in Einstein-like notation, is crucial for the continued adoption of tensors. We use the motivating example of high-order differential operators to illustrate this need. As sparse tensor arithmetic represents an emerging research topic, with challenges distinct from the \ac{MV} paradigm, many aspects require further articulation and development.  This work focuses on three core facets. First, aligning with prominent voices in the field, we emphasise the importance of data structures able to accommodate the operational complexity of tensor arithmetic. However, we describe a \ac{LCO} data structure that provides faster and more memory-efficient sorting performance in support of this operational complexity. Second, flexible data structures, like the \ac{LCO}, rely heavily on sorts and permutations. We introduce an innovative permutation algorithm, based on radix sort, that is tailored to rearrange already-sorted sparse data, producing significant performance gains. Third, we introduce a novel poly-algorithm for sparse tensor products, where hyper-sparsity is a possibility. Different manifestations of hyper-sparsity demand their own customised approach, which our multiplication poly-algorithm is the first to provide. These developments are incorporated within our LibNT and NTToolbox software libraries. Benchmarks, frequently drawn from the high-order differential operators example, demonstrate the practical impact of our routines, with speed-ups of $40\%$ or higher compared to alternative high-performance implementations. Comparisons against the MATLAB Tensor Toolbox show over $10$ times speed improvements.  Thus, these advancements produce significant practical improvements for sparse tensor arithmetic.

\end{abstract}



\pagestyle{myheadings}
\thispagestyle{plain}
\markboth{A.~P. HARRISON AND D. JOSEPH}{HIGH PERF ROUTINES FOR SPARSE TENSOR ARITH}

\acresetall

\section{Introduction}

Efficient sparse-matrix data structures and algorithms have had an unquestionably massive impact on the field of technical computing. In recent years, however, algebra and calculations for high-order\footnote{We use order to denote the number of modes or indices of a tensor.} data, which we call \acp{NT}~\cite{Harrison_2016_ten1}, have begun to enjoy a prominent role. We use the ``numeric'' adjective to differentiate this research topic from \emph{geometric} tensors. The latter must meet strict mathematical or physical properties, whereas this is not necessarily the case for all uses of numeric high-order data. Thus, for the remainder of this work, we use ``tensor'' to mean the numeric variant.


In fact, the need for mature sparse computations may be even more acute in the tensor domain than in the \ac{MV} domain. If using a dense representation, the computational and memory demands of working with tensors grows exponentially with order, raising barriers to tractability. For this reason, circumventing ``the curse of dimensionality'' that harries tensor operations is a crucial goal. Apart from work into low-parametric representations~\cite{Beylkin_2005,Oseledets_2009,Oseledets_2012,Khoromskij_2012} and implicit representations of high-order operators~\cite{Ahlander_2002,Ahlander_2006,Cummings_1999}, researchers have also focused on how to explicitly represent sparse tensors~\cite{Chang_2001,Lin_2002,Lin_2003,Bader_2007,Gundersen_2012,Baskaran_2012,Kang_2012,Parkhill_2010}.

Tensor decomposition drives a great deal of this work~\cite{Chang_2001,Lin_2002,Lin_2003,Bader_2007,Baskaran_2012,Kang_2012}, but applications involving high-order linear operators~\cite{Ahlander_2002,Ahlander_2006,Cummings_1999,Parkhill_2010},  high-order partial derivatives~\cite{Gundersen_2012}, and deep learning~\cite{pytorch} also see need for sparse tensors. Additionally, the known link between symmetric tensors and polynomial equations~\cite{Comon_2008} introduces further impetus for sparse tensor computations. Considering the high-sparsity of real-world polynomial equations~\cite{Stetter_2004}, if tensors are used to manipulate such equations, efficient sparse algorithms will be needed (likely along with symmetric-specific optimizations~\cite{Schatz_2014}).

As Bader and Kolda~\cite{Bader_2006,Bader_2007} and we~\cite{Harrison_2016_ten1} have argued, a suite of core tensor data structures and algorithms can allow fast prototyping and development of algorithms applied to high-order data. A further argument can be made that multi-purpose algorithms and data structures for tensor \emph{arithmetic} operations, \ie{} those found in Einstein-like notation~\cite{Harrison_2016_ten1,Harshman_2001,Ahlander_2002,Ahlander_2006}, are a valuable tool for working with and understanding tensor operations~\cite{Harrison_2016_ten1,Ahlander_2002,Ahlander_2006}. These include multiplication, solution of linear equations, and addition/subtraction. Combined with an interface that supports Einstein-like notation, core tensor arithmetic computations would be an important part of a technical computing framework for tensors, analogous to the environments used for \ac{MV} algebra and computations, \eg{} MATLAB. We note that core arithmetic routines would not obviate the need for specialised data structures and algorithms, \eg{} those used for the ubiquitous \ac{ALS} algorithm~\cite{Kolda_2009}.



The topic of data structures and algorithms for multi-purpose sparse tensor arithmetic has been broached previously by Bader and Kolda~\cite{Bader_2007} as part of their \ac{MTT}. Yet, since the \ac{MTT} does not provide high-performance kernels of its own~\cite{Schatz_2014}, there is considerable opportunity for continued investigation. With this vision in mind, this work offers a set of core kernels for sparse tensor arithmetic. Sharing Bader and Kolda's~\cite{Bader_2007} design philosophy of not favouring any particular index over another, this work describes a \ac{LCO} sparse-tensor data structure, which is related to, but different from, the one seen in the \ac{MTT}. The flexibility and simplicity of the \ac{LCO} data structure comes at the cost of heavily relying on sorts and permutes. This work describes high-performance rearrangement algorithms specifically tailored for sparse tensors. Finally, this paper describes a multiplication poly-algorithm that can effectively compute the products between any tensors exhibiting any manner of sparsity, including hyper-sparsity.

Detailed benchmarks demonstrate the high performance of these algorithms. To provide a motivating example, many of the benchmarks are drawn from the exemplar of using Einstein-like notation to construct high-order differential operators. Because the impact of different data-structure and algorithmic choices are just beginning to be understood within sparse tensor arithmetic, we limit our focus to sequential implementations. This also makes any performance comparisons with the \ac{MTT} more fair, whose sparse tensor arithmetic functionality is predominantly based on sequential algorithms within MATLAB. All computations are implemented within our open-source LibNT and NTToolbox software libraries\footnote{https://github.com/extragoya/LibNT}, whose dense tensor routines have been previously introduced as part of the \ac{NT} framework~\cite{Harrison_2016_ten1}.

\Sec\ref{sec:prelim} begins by using the example of high-order differential operators, applied to images, to motivate the development of high-performance routines to support the tensor arithmetic operations found in Einstein-like notation. With these preliminaries discussed, \Sec\ref{sec:data_representation} outlines a flexible data representation for sparse tensors. This data representation places a heavy burden on fast methods to rearrange data, which \Sec\ref{sec:sort_permute} addresses by outlining algorithms to permute sparse tensors. Sparse-tensor multiplication is discussed in \Sec\ref{sec:sparse_mult}. Comparative performance with the \ac{MTT}, and a high-performance re-implementation of its multiplication strategy, is highlighted in \Sec\ref{sec:sparse_results}. Finally \Sec\ref{sec:sparse_conclusion} discusses and concludes this work. Tests were performed on a Windows $64$-bit workstation, using an Intel E8400 CPU with 8\,Gb of memory. All algorithms are implemented as part of LibNT's C++ code, to which the MATLAB library NTToolbox interfaces.

\section{High-Order Differential Operators}
\label{sec:prelim}

Many scientific domains require software tools to work with and algebraically manipulate high-order numeric data~\cite{Harrison_2016_ten1}. High-order differential operators are one such important exemplar~\cite{Ahlander_2002,Ahlander_2006}. We make this more concrete by focusing on high-order operators applied in computer vision using an Einstein-like notation, but we emphasise that the need for sparse tensor arithmetic transcends both computer vision and high-order differential operators.

Applied to gridded data, \eg{} an image, often representing a partial differential equation, differential operators commonly take the form of \ac{FD} operators. Such operators may be explicitly needed within optimization problems, \eg{} where the operand of the differential operator is an unknown that must be solved using a least squares method. For length $N$ first-order data, sparse \ac{FD} matrix operators are relatively easy to construct. For example, should $O(h^{2})$ central differencing be required, the sparse \ac{FD} matrix can be constructed with \ac{MV} algebra using
\begin{align}
    \textbf{D}=\dfrac{1}{2}
            \left\{\left(
                 \begin{array}{ccc}
                   \textbf{0} & 0 & 0 \\
                   -\textbf{I} & \textbf{0} & \textbf{0} \\
                   \textbf{0} & 0 & 0 \\
                 \end{array}
               \right) +
               \left(
                 \begin{array}{ccc}
                   0 & 0 & \textbf{0} \\
                   \textbf{0} & \textbf{0} & \textbf{I}  \\
                   0 & 0 & \textbf{0} \\
                 \end{array}
               \right)
               +
               \left(
                 \begin{array}{ccccccc}
                   -3 & 4 & -1 & \textbf{0} & 0 & 0 & 0\\
                   \textbf{0} & \textbf{0} & \textbf{0} & \textbf{0} & \textbf{0} & \textbf{0} & \textbf{0}\\
                   0 & 0 & 0 & \textbf{0} & -1 & 4 & -3\\
                 \end{array}
               \right) \right\}
               \textrm{,} \label{eqn:FD_matrix}
\end{align}
where the first and last rows of $\textbf{D}$ are filled using $O(h^{2})$ forward and backward-differencing operators. Each term in \eqref{eqn:FD_matrix} is an $N\times N$ matrix, meaning $\mathbf{I}$ is of size $(N-2)\times (N-2)$. The $\textbf{0}$ sub-matrices are sized based on this and whether they share a row or column with $\mathbf{I}$ or a scalar. Unfortunately, working with \ac{FD} operators of higher order using \ac{MV} algebra and software can be prohibitively challenging and error prone, as users must work with ``flattened'' versions of the operators and operands~\cite{Ahlander_2002,Ahlander_2006}. The challenges only multiply as  order increases, \eg{} 3D medical imaging scans, or a time series of such scans. Thus, it can be beneficial to express and programmatically construct such operators within their natural high-order domain~\cite{Ahlander_2002,Ahlander_2006}.

\begin{figure}
  \centering
  \includegraphics[scale=.5]{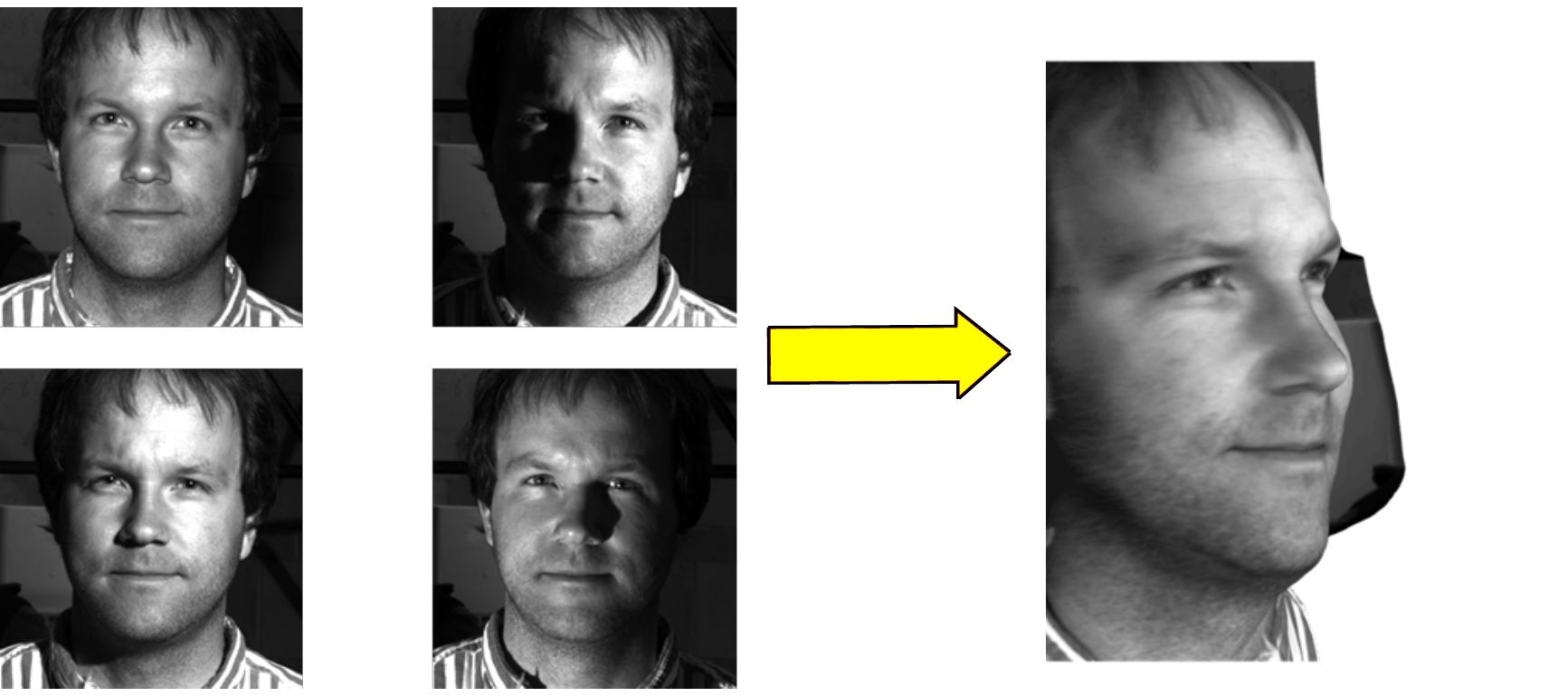}
  \mycaption{Illustration of depth-map \& albedo estimation.}{Four images of a single-view image sequence are displayed. Each image corresponds to a distinct principal light direction, with its own shading characteristics. From the image sequence, an estimate of the depth of the subject's face, along with the albedo, can be produced through a combinatorial Laplacian formulation. Although not discussed here, anisotropic elements, using entry-wise products, are frequently required to manage image noise~\cite{Harrison_2012b}. Image data obtained from the Extended Yale Face Database B~\cite{Georghiades_2001}.}\label{fig:depth_map_estimation}
\end{figure}

When dealing with generative models used in computer vision, often \ac{FD} operators across different image indices are incorporated. One of many examples is depth-map and albedo estimation~\cite{Harrison_2012b}, illustrated by \Fig~\ref{fig:depth_map_estimation}.
Estimating the phenomenon in question requires constructing a design tensor and inverting a sparse system of high-order linear equations. Using Einstein-like notation~\cite{Harrison_2016_ten1,Harshman_2001,Ahlander_2002,Ahlander_2006}, and assuming a 2D image domain, such systems can be expressed in general terms as
\begin{align}
    y_{ij}=a_{ijk\ell}x_{k\ell} \textrm{,} \label{eqn:laplace_system}
\end{align}
where repeated and non-repeated indices denote inner and outer products, respectively, $x_{k\ell}$ is the tensor of quantities of interest, \eg{} a depth-map, $y_{ij}$ are observations relevant to the generative model, and $a_{ijk\ell}$ is a sparse design tensor. A common makeup of the design tensor is the combinatorial Laplacian:
\begin{align}
    a_{ijk\ell}=d^{(y)}_{ii'}\delta_{j\ell}d^{(y)}_{i'k}+d^{(x)}_{jj'}\delta_{ik}d^{(x)}_{j'\ell} \textrm{,} \label{eqn:isotropic_new}
\end{align}
where $d^{(.)}_{ij}$ is the sparse tensor version of the \ac{FD} matrix of \eqref{eqn:FD_matrix} and $\delta_{ij}$ is the Kronecker delta. Frequently, the Laplacian operator is also anisotropic, \eg{} to manage image noise~\cite{Harrison_2012b}, and nonlinear formulations can also be required~\cite{Harrison_2015_SPIE,Harrison_2015_phd}. Representing such complexities benefits further from Einstein-like notation, but for the sake of simplicity we limit the exposition to the isotropic and linear example of \eqref{eqn:isotropic_new}.

Equipped with an appropriate computational environment, a researcher or practitioner could programmatically construct $a_{ijk\ell}$ using an Einstein-like notation. For example, LibNT and/or NTToolbox~\cite{Harrison_2016_ten1} will execute \eqref{eqn:isotropic_new} as binary operations left-to-right with operator precedence\footnote{This follows the same practice found in numerical packages like MATLAB and NumPy. Although LibNT's C++ interface uses rvalue references and pools of memory, temporary memory reallocation still occurs. NTToolbox's MATLAB interface does not have the same level of optimisation.}. As we will elaborate, the executed products include hyper-sparse matrix products, requiring a strategy different from those found in standard sparse matrix multiplication. In addition,  the multiplication, addition, and assignment operations will require rearrangements of non-zero values. Thus, there is a demand here for sparse tensor rearrangement and multiplication algorithms. We will return to these equations to provide an application-based context for this work's contributions.

\section{Data Representation}
\label{sec:data_representation}

This section first outlines considerations for sparse tensor representation, highlighting the \ac{LCO} format used in LibNT. Afterwards, some comparative results demonstrate the \ac{LCO} format's effectiveness.

\subsection{Concepts}

Sparse \ac{MV} computations rely on compressed formats~\cite{Davis_2006,Buluc_2011b}, \eg{} the \ac{CSC} format, which allows efficient column-centred operations at the expense of inefficient row-centred operations. However, Bader and Kolda~\cite{Bader_2007} convincingly argue against compressed tensor formats, as it requires categorising an index, or set of indices, differently from others, which becomes less meaningful as the tensor order increases.

These arguments are bolstered by considering tensor arithmetic's operational complexity, \ie{} the enormous number of ways that tensor indices can match up differently for the same arithmetic operation. For instance, enumerating all possible inner/outer product possibilities between two $N$-order tensors requires calculating all \emph{partial} permutations of the $N$ indices. Partial permutations~\cite{Bender_1991} are calculated using,
\begin{align}
    P=\sum_{i=0}^{N}i!{N \choose i}^{2} \,\textrm{,}
\end{align}
which grows factorially with order.
Similar conclusions are drawn when considering the index matchings within Einstein-like notation for addition and subtraction, as the number of possible index matchings also increases factorially with order.

This operational complexity suggests that data structures for sparse tensor arithmetic should be as flexible as possible. While several authors have adapted the compressed approach for non-arithmetic purposes~\cite{Chang_2001,Lin_2002,Lin_2003,Gundersen_2012}, these solutions would struggle to accommodate arithmetic operational complexity. Compression schemes would need to be re-computed or there would have to be different code implementations depending on what tasks are performed on what indices. Such strategies become less viable for tensor arithmetic with each increase in order.

Bader and Kolda make the case for a concurrent list of non-zero data and index \emph{values}. When an operation demands a different lexicographical order a rearrangement, \ie{} a sort or permute, is required. Thus, no indices are favoured over others in terms of operational efficiency, but rearrangements then play a heavy role. This approach has also been used within computational chemistry~\cite{Parkhill_2010} and deep learning~\cite{pytorch}.

The \ac{CO} and \ac{LCO} sparse formats are the two main non-compressed choices that store their non-zeros using straightforward lists. The \ac{CO} format stores expanded index values, \ie{} for an $N$-order tensor each of the $N$ index values. In contrast, the \ac{LCO} format stores \acp{LIV}, \ie{} $N$ index values represented by a single integer value. For instance, the zero-based \acp{LIV} for a third-order tensor, $a_{ijk}$, can be calculated using
\begin{align}
    \LIV=i+n_{i}(j+n_{j}k) \textrm{,}
\end{align}
where $n_{(.)}$ denotes the range of the corresponding index. Such a lexicographical order places greatest significance on the third index, followed by the second and first indices, which we designate numerically as $\{0,1,2\}$. Any permutation of the $\{0,1,2\}$ sequence is also valid, and this scheme is trivially extended to higher orders. \Table~\ref{tbl:sparse_formats} illustrates the differences between the two formats. Of note is that the sparse formats are identical for first-order tensors.
\begin{table}
\centering
\footnotesize
\mycaption{The \ac{CO} and \ac{LCO} sparse formats.}{Example zero-based index values from a $4\times4\times4$ sparse tensor, with a lexicographical order of $\{0,1,2\}$, illustrate the two formats. }
  \begin{tabular}{|c|c|c|c|c|c|c|}
  \hline
    \ac{CO} Index Values: & $\{1,0,0\}$ & $\{2,0,1\}$ & $\{0,1,1\}$ & $\{3,2,2\}$ & $\{1,0,3\}$ & $\{2,2,3\}$\\
    \hline
    \ac{LCO} Index Values: & $1$ & $18$ & $20$ & $43$ & $49$ & $58$\\
  \hline
  \end{tabular}
  \label{tbl:sparse_formats}
\end{table}

Both formats rely on a lexicographical order to arrange non-zero values. For the \ac{CO} format, $\{0,1,2\}$ indicates that when comparing values, the third index value must be considered first, followed by the second and first index values. Altering the lexicographical order requires changing the sequence in which expanded index values are compared. In contrast, for the \ac{LCO} format, the lexicographical order governs the linearisation scheme used to compute \acp{LIV}. Changing the lexicographical order necessitates recomputing \acp{LIV}, which we call \ac{LIV} \emph{shuffle}. Once done, a straightforward integer comparison then suffices to compare \acp{LIV}.

Bader and Kolda opt for the \ac{CO} sparse format for the \ac{MTT} library~\cite{Bader_2007}. While Bader and Kolda do not specifically discuss the \ac{LCO} format, they do mention concerns with linearisation schemes in general, arguing that \acp{LIV} may overflow integer datatypes. This is a valid concern. However, many applications, such as computer vision~\cite{Harrison_2015_phd}, computational chemistry~\cite{Parkhill_2010}, and deep learning~\cite{pytorch}, often employ tensors whose dimensionalities fit within a 64-bit limit (or 63-bit limit if using signed integers). For cases where \acp{LIV} do exceed standard integer limits, \eg{} problems involving the SNAP dataset~\cite{McAuley_2013}, high-precision integer libraries~\cite{Boost_multi,GMP} could offer very-large \acp{LIV}. Nonetheless, here we limit our scope to tensors whose dimensionality fits within the signed integer limits of $2^{63}-1$, leaving the topic of very-large \acp{LIV} to future work.

Moving on from overflow issues, other factors also play an important role. For instance, compared to the \ac{CO} format and assuming all index values are stored using the same fixed-sized integer datatype, the \ac{LCO} format is more memory efficient for tensor orders greater than one. With additional bookkeeping, less memory could be used in the \ac{CO} format by employing variable-sized integers, \eg{} choosing 8-bit, 16-bit, 32-bit, or 64-bit integers for each index based on its dimension. However, such extra bookkeeping and complexity comes with its own penalties. Moreover, developing an efficient implementation using high-performance statically-typed languages, without using costly dynamic polymorphistic operations, is not easily resolved. As a result, the exploration of variable-sized \ac{CO} indices is left for future work.

With the above caveats in mind, there is an increased cost of certain fundamental operations when using the \ac{CO} format. For instance, comparison operations in the \ac{CO} format require up to $N$ individual numerical comparisons for an $N$-order tensor. Such comparison operations are fundamental kernels within sorting algorithms and arithmetic operations built on the format. Additionally, the increased memory requirements degrade locality between consecutive non-zero index values, resulting in more cache misses, which can be the deciding factor in sorting performance~\cite{LaMarca_1999}. This also impacts arithmetic operations. These considerations all add up to the \ac{CO} format placing greater demands on memory bandwidth, which is often the limiting factor in modern computer architectures~\cite{Drepper_2007}.

On the other hand, the \ac{LCO} format requires an \ac{LIV} shuffle to change lexicographical orders. Thus, putting memory storage requirements aside, choosing between the two can come down to comparing the impact of the increased comparison, read, and write \ac{CO} costs vs. the $O(\nnz)$ \ac{LIV} shuffle step of the \ac{LCO} format. Benchmark tests can measure the relative impact of these costs.

\subsection{Benchmarks}
\label{sec:data_format_test}
As the example in \Sec\ref{sec:prelim} highlights, rearrangements of non-zero data is a frequent requirement for tensor arithmetic. For instance, to perform the addition in \eqref{eqn:isotropic_new}, one of the terms must be re-sorted based on the index matching. We call rearranging already sorted data into a new lexicographical order \emph{permutation}, which can benefit from specialised algorithms that we discuss in more detail in \Sec\ref{sec:sort_permute}. However, for the sake of simplicity, here we focus on \emph{sorting} algorithms to compare the two formats. Sorting is typically required when non-zero data is unsorted, \eg{} after tensor construction or the insertion of un-ordered non-zeros. We first discuss details on the data formats we test, followed by an explanation of the sorting algorithms used in the benchmarks. Afterwards we highlight the results of two different tests.

Two variants of the \ac{CO} format were tested. The first, denoted \ac{CO}\_Separate and used within the \ac{MTT}, uses contiguous memory regions to store specific expanded index values, \eg{} the first coordinates are stored contiguously, followed by the second, and so on. We also tested a second variant that packs expanded index values consecutively one after each other, thereby better maximising memory locality across consecutive accesses of tensor elements. We call this variant \ac{CO}\_Packed. If the $N$ index values of each of the $M$ non-zeros were stored in an $M\times N$ matrix, \ac{CO}\_Separate and  \ac{CO}\_Packed would arrange them in column- and row-major order, respectively.

As changing the lexicographical order is often performed prior to rearranging data, we also measure the \ac{LIV} shuffle cost for the \ac{LCO} format. If executed naively, this operation can be very expensive as \ac{LIV} shuffles require integer division. However, we use a fast division library~\cite{libdivide} to mitigate this cost.

For the most part, experiments are restricted to using signed $64$-bit integers to store index values\footnote{For software engineering reasons we use signed integers to avoid undefined behaviour if an index is decremented beyond zero.}. Nonetheless, we do briefly explore the use of \emph{fixed} $16$-bit signed integers for the \ac{CO} format to help shed light on any performance impacts of using smaller-sized integers.

Tests employed two well-known sorting algorithms. The first corresponds to the introspective sorting algorithm~\cite{Musser_1997}, used in the C++ standard and considered a gold-standard~\cite{Meyers_2001}. The second corresponds to \ac{MSD} radix sort~\cite{Sedgewick_1998}, which, unlike general sorting algorithms, is designed specifically for integer-like data. Experiments used C++ implementations, adapted from optimised and publicly available general-purpose versions~\cite{Schwarz_2006,MSD_radix_inplace} to handle the \ac{CO}\_Separate, \ac{CO}\_Packed, and the \ac{LCO} formats, along with the accompanying data array. Attesting to their speed, we found that in our tests our \ac{CO}\_Separate implementation always outperformed MATLAB's \texttt{sortrows}, which is the approach the \ac{MTT} uses to sort its \ac{CO}\_Separate data. Code can be found within our publicly available LibNT library. More details on our implementations can be found in our supplemental material.

The first test measured times to sort a fifth-order sparse tensor. As mentioned, we also recorded the time needed to shuffle \ac{LIV} values. As integer division operations are extraordinarily fast when divisors are a power of two, index ranges were chosen to be $2^{10}-1$ to avoid providing the \ac{LCO} format with an unfair advantage. To judge the impact of tensor order, the same tensor was also ``flattened'' into lower-order \ac{LCO} and \ac{CO} formats.  Thus the impact of increasing tensor order, with its increased demands on memory bandwidth and \ac{LIV} shuffles, was measured under identical conditions.

\begin{figure}
  \setlength{\tabcolsep}{1pt} 

  \center
  \includegraphics{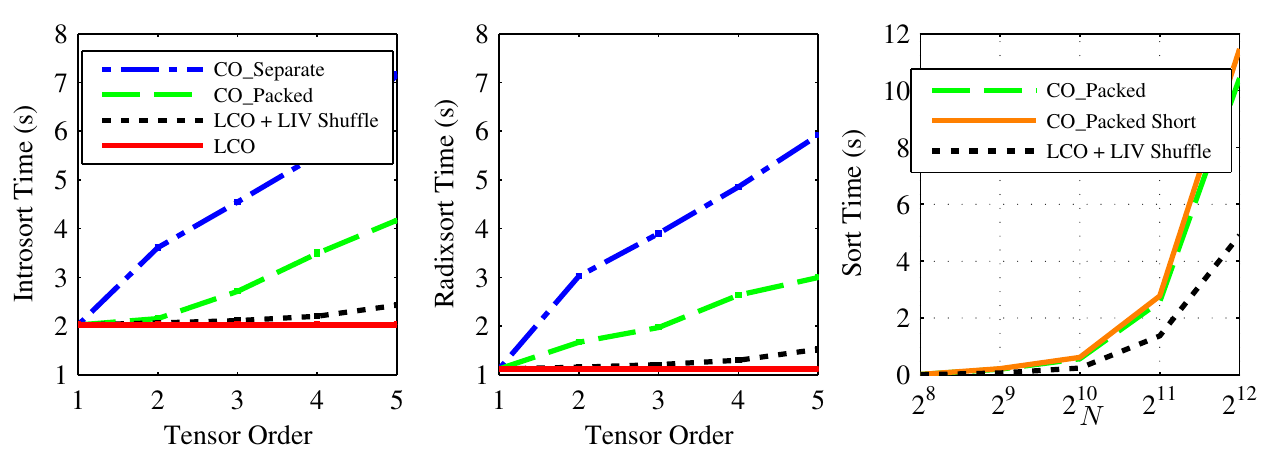}
  \begin{tabularx}{\textwidth}{YYY}
  (a) & (b) & (c)
  \end{tabularx}
  \caption{Sorting times of the \ac{LCO}, \ac{CO}\_Packed and \ac{CO}\_Separate sparse formats. (a) and (b) depict the time to sort a fifth-order $N \times N \times N \times N \times N$ tensor, with $N=2^{10}-1$ and $5N^{2}$ non-zeros, using the introspective and \ac{MSD} radix sort algorithms, respectively. The time taken with flattened lesser-order versions and with shuffling \acp{LIV} was also measured. (c) depicts the \ac{MSD} radix sort run time of \ac{CO}\_Packed Short, a $16$-bit variant of \ac{CO}\_Packed,  and \ac{LCO} plus an \ac{LIV} shuffle to perform the sort needed to add the fourth-order tensors in \eqref{eqn:isotropic_new}.}
\label{fig:co_vs_dok_table}
\end{figure}

\Fig{}s~\ref{fig:co_vs_dok_table}(a) and (b) outline the results of this first test, using introspective and \ac{MSD} radix sort, respectively.
As the figures demonstrate, shuffling \acp{LIV} comes with a non-trivial running-time cost, which increases with order. However, the cost of sorting both variants of the \ac{CO} format is greater, meaning that even with an \ac{LIV} shuffle included, sorting \ac{LCO} indices is still much faster than sorting second-order or higher \ac{CO} indices.

To contextualise these results within the differential operators application, we also measured sorting times to add the two fourth-order tensor terms in \eqref{eqn:isotropic_new} using the best performing algorithm  of \ac{MSD} radix sort applied to the \ac{LCO} and CO\_Packed formats. These terms are formed after the necessary products are executed, and can be denoted $c_{ij\ell k}^{(y)}$ and $c_{jik\ell }^{(x)}$, where the ordering of indices post-multiplication follows LibNT's conventions~\cite{Harrison_2015_phd}. We assume both tensors lie in the $\{0,1,2,3\}$ lexicographical order, meaning to perform the addition one of the tensors must be re-sorted into the $\{1,0,3,2\}$ lexicographical order. While this is technically a permutation task, measuring differences in using general-purpose sorting algorithms can still be informative to assess data-format performances. We assume input images are of size $N\times N$, making each tensor size $N\times N \times N \times N$, and we measure sorting times for increasing values of $N$. \Fig~\ref{fig:co_vs_dok_table}(c) depicts a plot of the sorting times, demonstrating the large gain in speed of the \ac{LCO} format. In particular, at the highest value of $N$, the \ac{LCO} format plus the \ac{LIV} shuffle consumes $4.9\,s$, whereas \ac{CO}\_Packed consumes $10.4\,s$. 

For the highest $N$, once the tensors are sorted the run time for the addition operation is only $2.7\,s$ for the \ac{LCO} format. Thus, sorting the \ac{LCO} format takes roughly $65\%$ of the total operation time. In contrast, for the \ac{CO} format, should the addition operation consume roughly the same amount of time, sorting would consume $80\%$ of the total time, demonstrating both the importance of optimising rearrangements and the importance of data format choice for sparse tensor arithmetic. 

To shed light on the prospect of using variable-sized integers, we also tested the sorting performance of CO\_Packed when using \emph{fixed} $16$ bit integers to hold each coordinate. While this scheme does not address how to best implement a variable-sized approach, it does help reveal if using smaller-sized integers may allow CO\_Packed to outperform LCO. However, as \Fig~\ref{fig:co_vs_dok_table}(c) demonstrates, the $16$-bit variant of CO\_Packed consistently ran slightly slower than the $64$-bit variant. One possible explanation for this is that $64$-bit architectures, like the one used for testing, may be better optimised to operate in its native bit size. While this question does deserve further investigation, these preliminary tests further support the conclusion that \ac{LCO} enables faster sorting speed regardless of the underlying \ac{CO} datatype.


In sum, these results indicate that the \ac{LCO} format is better able to manage the demands of increasing tensor orders. This is crucial when operating with sparse tensors of high orders, such as those seen in computational chemistry~\cite{Parkhill_2010,Kats_2013}, computer vision~\cite{Harrison_2015_phd} or deep learning~\cite{pytorch}. Considerable performance differences were also evident at lower orders. Coupled with the fact that the \ac{LCO} format uses much less memory at high orders when using native bit sizes, these performance metrics lead us to prefer the \ac{LCO} sparse format over either \ac{CO} format variant.

\section{Permutation}
\label{sec:sort_permute}

As noted, a non-compressed sparse format places a heavy demand on rearranging non-zero data. Consequently, fast and efficient sparse tensor arithmetic can hinge on the algorithmic choices made for rearrangements. For sorting, this was demonstrated by \Fig~\ref{fig:co_vs_dok_table}(a) vs (b), where \ac{MSD} radix sort performed roughly twice as fast as introsort. For this reason, we opt for \ac{MSD} radix sort as the sorting algorithm for sparse tensors. Our supplemental material includes more extensive experiments supporting this conclusion, comparing \ac{MSD} radix sort against three other leading algorithms.

Yet, permutation, \ie{} rearranging already sorted data into a different lexicographical order, is arguably even more important than sorting. In the \ac{MV} paradigm such tasks are called transposition. Because of operational complexity, tensors are frequently arranged in an undesired lexicographical order, making sparse permutation a frequent first step within tensor arithmetic. In fact, this was already demonstrated in the benchmarks of \Fig~\ref{fig:co_vs_dok_table}(c), where a permutation was required to perform the sum in \eqref{eqn:isotropic_new}. While permutation operations are tasked with the same goal as sorting, \ie{} rearranging data into a desired lexicographical order, their starting points differ. By taking advantage of the existing structure of already sorted non-zero data, faster means to permutation can be realised. These speedups can be quantified using benchmarks.

\subsection{Algorithm}

When permuting data the first step is to recompute the \acp{LIV} into the new lexicographical order. The work needed for the subsequent rearrangement depends on the relationship between the starting and ending lexicographical orders. For instance, intuitively it should be simpler to permute sparse tensor data from the $\{0,1,2,3\}$ lexicographical order to the $\{1,0,2,3\}$ lexicographical order than it would be to permute it to the $\{3,2,1,0\}$ lexicographical order. The former only rearranges two indices, while the latter rearranges all of them. This intuition stems from the fact that regardless of their starting and ending lexicographical orders, new \acp{LIV} will always be arranged in sequences of \emph{sorted sublists}. Specific subsequences of these sublists must be merged together, creating new sequences of sorted lists that may be in the right arrangement or may require additional merges. This relationship can be formalised, providing for a ready identification of efficiencies.

\begin{figure}
  \center
  \includegraphics[scale=1]{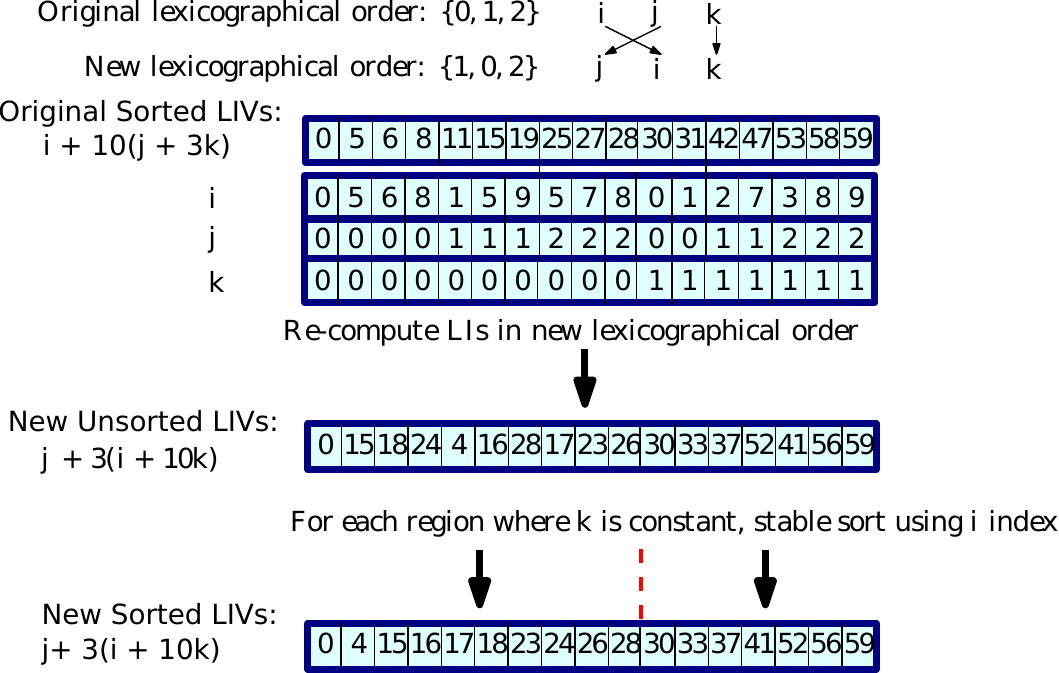}
  \mycaption{Permuting a $10\times 3 \times 2 $ third-order tensor, $a_{ijk}$.}{The figure depicts the original and new \acp{LIV} corresponding to starting and ending lexicographical orders of $\{0,1,2\}$ and $\{1,0,2\}$, respectively. The index of highest significance in the new \acp{LIV} is a resting index, so regions where $k$ is constant can be sorted independently. Each such independent region must be stably sorted based on the rearrangement index $i$. The final resting index $j$ can be ignored. Note that the $i$, $j$, and $k$ indices are rendered for the sake of illustration, but in the \ac{RP} algorithm indices are computed on-the-fly using integer division and modulo operations.  }
  \label{fig:complete_permutation}
\end{figure}

A permutation essentially divides tensor indices into two sets---those that require rearranging and those that do not. \Fig~\ref{fig:complete_permutation} illustrates how this can be determined, with a third-order tensor $a_{ijk}$.
The top of the figure illustrates the bipartite graph of the starting and ending arrangements of $i$,$j$,$k$. Because the $i$ index crosses an index that \emph{originally had} a higher significance, \ie{} $j$, the new \acp{LIV} must be rearranged according to the $i$ index. We call such indices \emph{rearrangement indices}. The other two indices do not meet this criterion and thus, the new \acp{LIV} do not need to be rearranged according to $j$ and $k$. These indices we call \emph{resting indices}.

Categorising the indices this way breaks the permutation task into a recursive hierarchy of steps. For instance, working from highest-to-lowest significance of the \emph{new} \acp{LIV} in \Fig~\ref{fig:complete_permutation}'s example, $k$ is a resting index. As a result, regions where $k$ is constant can each be independently sorted. These independent regions can be \emph{stably} sorted based solely on the $i$ index, which is a rearrangement index. Stability means the original relative ordering is used to break ties between equal values. The next resting index $j$ is also the final index, so there is no more work to do. However, if $j$ was not the final index, then the process would have to continue, where each sub-region where $j$ is constant would be sorted. This process can be generalised to arbitrary orders and starting/ending lexicographical orders. An important aspect to note is that the final index is always a resting index.

Returning to \Fig~\ref{fig:complete_permutation}'s example, identifying regions in the new \acp{LIV} where $k$ is constant can be done by integer dividing the starting \ac{LIV} by $n_{j}n_{i}=30$ to compute the $k$ value, and then computing the maximum possible \ac{LIV} at that value of $k$. A linear scan that stops when this threshold is broken identifies the end-point of the region. The process can be repeated for the next value of $k$. Separating the \ac{LIV} values into independent parts benefits all sorting algorithms. For comparison sorts, the asymptotic bounds may be lowered. However, for radix sorts, within each region of constant $k$, each \ac{LIV} can be examined modulo $30$, reducing the maximum possible integer magnitude to accommodate. Depending on the radix digit size, this can reduce the key length, thereby reducing the number of passes a radix sort need perform. Moreover, when sorting each independent region of constant $k$, the \acp{LIV} modulo $30$ need only be stably sorted using the rearrangement index $i$. Thus each \ac{LIV} modulo $30$ can be reduced even further by integer dividing by $n_{j}=3$. In the general case, this aggressive shaving off of irrelevant portions of the \acp{LIV} can drastically reduce the key length for radix sorts, significantly reducing the number of passes the sort must perform.

To take advantage of these characteristics, LibNT includes an algorithm, called \ac{RP}. Given a starting and ending lexicographical order, a preprocessing step determines which indices are rearrangement or resting indices. The \ac{RP} routine then employs a stable, but not inplace, variant of the \ac{MSD} radix sort algorithm. Since shaving off irrelevant portions of the \acp{LIV} relies on integer division, libdivide~\cite{libdivide} is used to to minimise slowdowns. Nonetheless, even when using a fast integer-division library, shaving off \acp{LIV} comes with a computational cost, which can only be justified if the number of radix sort passes can be reduced. This is typically the case when both the \ac{NNZ} and reductions in \ac{LIV} magnitude are large. For this reason, LibNT's \ac{RP} algorithm is adaptive and will switch to the standard \ac{MSD} radix sort based on an estimate of the reduction in pass numbers.

There are theoretically interesting implications of permuting sparse tensor data this way. As Sedgewick explains, radix sorts are often sublinear in the information content of the keys being sorted~\cite{Sedgewick_1998}, meaning they can often arrange data without examining every bit. However, this is only an average-case result based on random conditions. Yet, in the context of sparse-tensor permutations, by always having \emph{at least} one index a resting index, it is always possible to permute without examining every bit in the \acp{LIV}. Whether these theoretical gains translate to practical ones is a matter revealed by benchmarks.

\subsection{Results}

\begin{figure}[t]
  \center
  \includegraphics{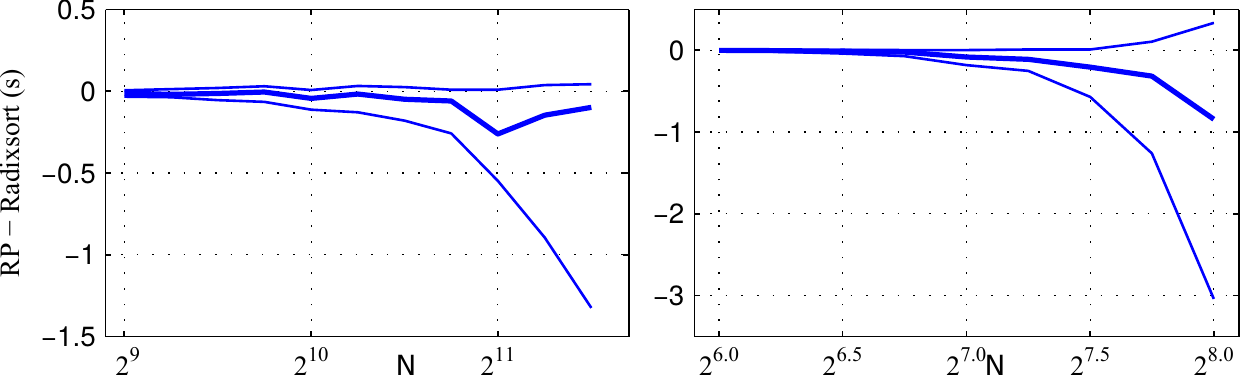}
  \begin{tabularx}{\textwidth}{YY}
    (a) & (b)
  \end{tabularx}
  \mycaption{Permutation benchmark results.}{(a) and (b) depict the differences in time between RP and radixsort to permute $N \times N \times N \times N$ tensors. Positive and negative values mean RP ran slower and faster, respectively. (a) and (b) use the combinatorial Laplacian in \eqref{eqn:isotropic_new} and a tensor with $2\%$ fill factor, respectively. All $23$ possible permutations were performed, with the heavy-weighted line displaying median run time differences across all permutations, and lighter-weighted lines depicting maximum and minimum differences.}
  \label{fig:permuting_results1}
\end{figure}

Tests measured the permutation performance of \ac{RP} using all $4!-1=23$ permutations of a fourth-order tensor. \Fig~\ref{fig:permuting_results1}(a) first depicts differences in run time between \ac{RP} and the \ac{MSD} radix sort algorithm (negative values are better for RP) to permute the fourth-order combinatorial Laplacian tensor in \eqref{eqn:isotropic_new}. This operator's fill factor decreases quadratically with dimensionality, which produces highly-sparse fill factors at large dimensionalities. \Fig~\ref{fig:permuting_results1}(b), on the other hand, depicts results of a fourth tensor whose fill factor remains a constant $2\%$. Both sets of tests were performed at increasing levels of dimensionality. While other algorithms were also tested, including those well suited to sorting already sorted sublists, \eg{} natural mergesort, only the \ac{MSD} radix sort algorithm proved competitive to \ac{RP}.

As the graph demonstrates, when examining the median run time across all permutations, \ac{RP} typically performed slightly better than \ac{MSD} radix sort, indicating that most permutations provide an opportunity for further optimization. More importantly, certain permutations provide even greater speed-up opportunities, with the \ac{RP} algorithm running at significantly faster speeds. To shed some more light on this, \Table~\ref{tbl:rp_timings} depicts timings corresponding to those of \Fig~\ref{fig:permuting_results1}(a) at $N=2^{11.5}$.
\begin{table}
  \centering
  \caption{Permutations and run times of \ac{RP} vs. \ac{MSD} radix sort, corresponding to the maximum, median, and minimum differences in time. All timings drawn from \Fig~\ref{fig:permuting_results1}(a) at $N=2^{11.5}$.}\label{tbl:rp_timings}
  \begin{tabular}{cccc}
     \hline
     Ranking & Permutation & \ac{MSD} Radix Sort (s) & \ac{RP} (s) \\
     \hline
     Max & $\{3,1,2,0\}$ & 3.50 & 3.54 \\
     Median & $\{2,1,3,0\}$ & 3.54 & 3.45 \\
     Min & $\{1,0,2,3\}$ & 2.71 & 1.38 \\
     \hline
   \end{tabular}
\end{table}
As can be seen, a permutation like $\{1,0,2,3\}$, which only requires that \ac{RP} rearrange according to the second index, allows a roughly $50\%$ increase in speed.

These results demonstrate that when opportunity affords, \ac{RP} can significantly speed up permutations. Considering that the \ac{MSD} radix sort already represents one of the fastest means to \emph{sort} \ac{LCO} data, these improvements attest to the value of using specialised \emph{permutation} algorithms. It is expected that these gains would only increase with higher orders and greater \acp{NNZ}.

\section{Multiplication}
\label{sec:sparse_mult}

Multiplying two sparse tensors together epitomises the unique demands of sparse tensor arithmetic. Any tensor multiplication, involving inner, entrywise, and outer products, can be represented as a sequence of matrix products~\cite{Harrison_2016_ten1}. Thus, sparse matrix products play a fundamental role in executing sparse tensor products. However, as \Sec\ref{sec:hyper} will explain, hyper-sparsity comes into play, calling for different strategies than those found in the \ac{MV} paradigm. Answering this need, \Sec\ref{sec:poly} outlines an effective multiplication poly-algorithm designed to handle operands exhibiting any manner of hyper-sparsity.

\subsection{Hyper-Sparsity}
\label{sec:hyper}

Sparse tensors often exhibit hyper-sparsity. Typically raised in an \ac{MV} context, hyper-sparsity refers to matrices where the numbers of rows and columns exceed the \ac{NNZ}~\cite{Buluc_2008,Buluc_2011}. Applied to a tensor context, this meaning implies a dimension  that exceeds the \ac{NNZ}. While comparatively rare in linear algebra, graph algorithm applications, which see uses for tensors~\cite{Dunlavy_2011}, encounter hyper-sparsity frequently~\cite{Buluc_2008,Buluc_2011}.

\begin{figure}[t]
  \center
  \begin{tabular}{ccc}
    \includegraphics{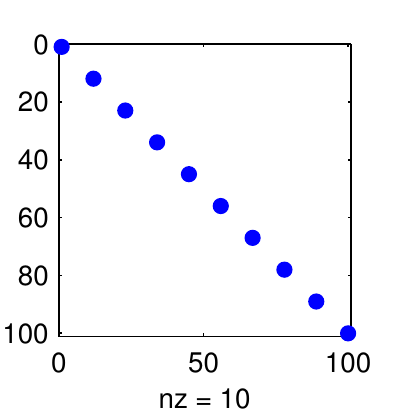} & \includegraphics[width=1.8in]{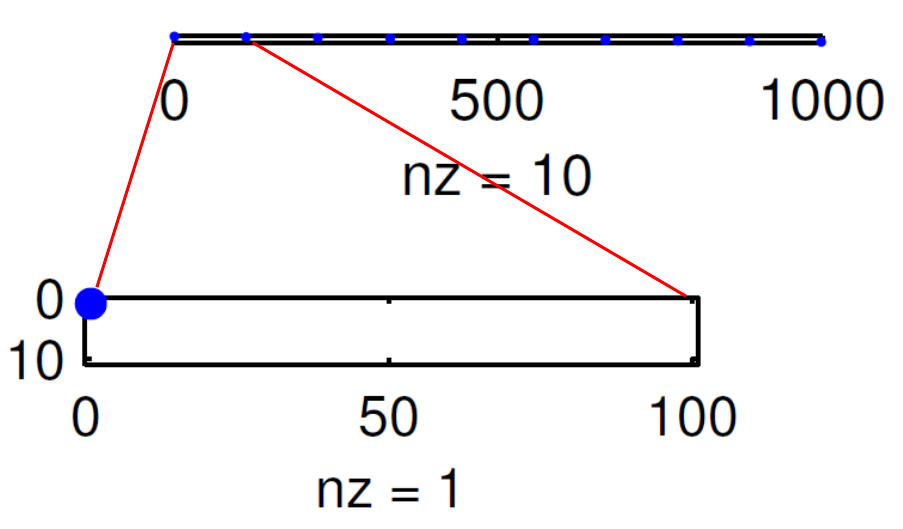} & \includegraphics[height=1.6in]{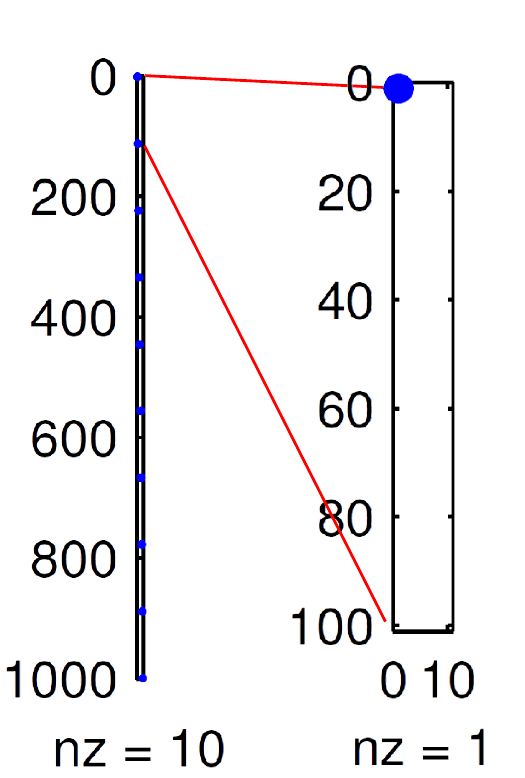} \\
    (a) & (b) & (c)
  \end{tabular}
  \mycaption{Hyper-sparsity of a matricised tensor.}{A fourth-order diagonal tensor $a_{ijk\ell}$, meaning all-zero except for when $i=j=k=\ell$, can produce hyper-sparse matrices when flattened: (a) depicts the hyper-sparsity pattern when a $10\times10\times10\times10$ diagonal tensor is matricised so that two of its indices are mapped to rows while the remaining are mapped to columns. An example multiplication causing this would be $a_{ijk\ell}a_{k\ell m n}$. (b) depicts the same tensor, except that it is matricised by mapping one index to rows, while the remaining are mapped to columns; (c) depicts the same as (b) except that the role of rows and columns are reversed. In the example multiplication of $a_{ijk\ell}a_{jk\ell m}$ the first and second operands would exhibit the patterns in (b) and (c), respectively. The two figures of (b) and (c) depict column- and row-sparsity, respectively, while (a) depicts index-sparsity. }
  \label{fig:hyper_sparse}
\end{figure}

However, even when tensors are not hyper-sparse on their own, they can exhibit hyper-sparsity when they are mapped to matrices during multiplication. For instance, as \Fig~\ref{fig:hyper_sparse}(a) demonstrates, flattening a purely diagonal tensor
can produce index-sparse matrices, meaning both row and column ranges exceed the \ac{NNZ}. In addition, as \Fig~\ref{fig:hyper_sparse}(b) and (c) demonstrate, flattening operations can also produce row- and column-sparse matrices, often manifesting as very-tall and very-wide matrices, respectively, in which only one of the dimensions exceeds the \ac{NNZ}. Such a case occurs in the differential operators example of \Sec\ref{sec:prelim}, which we return to in \Sec\ref{sec:sparse_results}. Thus, any routine executing sparse tensor products must be able to handle hyper-sparsity.

\subsection{Poly-Algorithm}
\label{sec:poly}

Sparse-tensor multiplication can be executed in three steps:
\begin{enumerate}
  \item map sparse-tensor \ac{LCO} data to a matrix, using a permute or sort if needed;
  \item convert each \ac{LCO} matrix to an appropriate \ac{MDT}, \eg{} \ac{CSC}, and multiply;
  \item map the resulting matrix back to a sparse tensor.
\end{enumerate}
This is similar to the scheme used by the \ac{MTT}. However, the \ac{MTT} always uses the \ac{CSC} \ac{MDT} and algorithm. However, the standard \ac{CSC} and \ac{CSR} \acp{MDT} are not equipped to handle hyper-sparsity~\cite{Buluc_2008,Buluc_2011}. This explains why  the \ac{MTT} excises all-zero columns and rows before performing standard \ac{CSC} multiplication, requiring that each matrix be sorted twice\footnote{In actuality the \ac{MTT} performs the excisions while the data is still in tensor form, but the implications for run time are identical.} and also necessitating additional bookkeeping which incurs its own running-time and memory costs. The reader is encouraged to consult our supplementary material for a more detailed explanation of the sorting and bookkeeping costs associated with excising all-zero rows and columns. In our experience, mirrored by others~\cite{Parkhill_2010} and in the tests of \Sec\ref{sec:data_format_test}, sorting or permuting costs can be a major run time cost in sparse tensor computations. For this reason, we minimise them as much as possible.

Thus, an attractive alternative is to employ algorithms and \acp{MDT} specialised to naturally handle the different types of hyper-sparsity that can possibly be encountered. Choosing different \acp{MDT} is not a freedom typically enjoyed within the \ac{MV} paradigm, in which matrices, once constructed, are typically locked into a single datatype and lexicographical order. Yet, this is not an issue in a sparse tensor context, where a conversion to an \ac{MDT} is required regardless. This extra flexibility calls for a poly-algorithm that dispatches to specific multiplication algorithms and \acp{MDT} depending on whether the flattened tensors are sparse, row-sparse, column-sparse, or index-sparse.

\Table~\ref{tbl:mult}
\begin{table}
\centering
\footnotesize
  \mycaption{Multiplication possibilities based on sparse characteristics of operands.}{Table entries indicate the algorithm LibNT employs, along with the section number describing it, for each sparse-characteristic combination.}
  \begin{tabular}{|C{.15\textwidth}|C{.15\textwidth}|C{.15\textwidth}|C{.18\textwidth}|C{.15\textwidth}|}
     \hline
     \rule{0pt}{4ex}  \backslashbox{$\mathbf{A}$}{$\mathbf{B}$} & Sparse & Row-Sparse & Column-Sparse & Index-Sparse \\
     \hline
       Sparse & \acs{CSC}/\acs{CSR} \newline (\Sec\ref{sec:sp_times_sp}) & \acs{CSC} \newline (\Sec\ref{sec:sp_times_sp}) & \acs{CSC} \newline (\Sec\ref{sec:sp_times_sp}) & \acs{CSC} \newline (\Sec\ref{sec:sp_times_sp}) \\
     \hline
      Column-Sparse & \acs{CSR} \newline (\Sec\ref{sec:sp_times_sp}) & \acs{DCSC}/\acs{DCSR} \newline (\Sec\ref{sec:dcsc}) & \acs{DCSC}\newline (\Sec\ref{sec:dcsc}) & \acs{DCSC} \newline (\Sec\ref{sec:dcsc}) \\
     \hline
      Row-Sparse & \acs{CSR} \newline (\Sec\ref{sec:sp_times_sp})& \acs{DCSR}\newline (\Sec\ref{sec:dcsc}) & \acs{CSCNA}/ \acs{CSRNA} \newline (\Sec\ref{sec:csc_no_accum}) & \acs{CSCNA} \newline (\Sec\ref{sec:csc_no_accum}) \\
     \hline
     Index-Sparse & \acs{CSR} \newline (\Sec\ref{sec:sp_times_sp}) & \acs{DCSR}\newline (\Sec\ref{sec:dcsc}) & \acs{CSRNA} \newline (\Sec\ref{sec:csc_no_accum}) & \acs{SOP} \newline (\Sec\ref{sec:outer_product}) \\
     \hline
   \end{tabular}
  \label{tbl:mult}
\end{table}
outlines the $16$ different possible sparsity combinations. In addition, it details the algorithmic choices used by LibNT. Two considerations motivated these choices. The primary consideration was on ensuring memory use and run time were not dependant on any hyper-sparse dimension sizes. As Bulu\c{c} and Gilbert~\cite{Buluc_2008,Buluc_2011} warn, algorithms, \eg{} the \ac{CSC}, whose run time and memory use depend on the dimensionalities of the matrix can consume inordinate amounts of memory or exhibit impractical run times under hyper-sparse conditions. With the first consideration satisfied, the second goal was gaining the fastest run time and/or the lowest memory use. Unlike \ac{MV} computations, performance metrics of sparse-tensor multiplication must include the cost of converting to the \ac{MDT}.

In describing the different multiplication algorithms, this subsection will use a set of common notation outlined in \Table~\ref{tbl:mult_terms}. While entrywise products are an important concept in tensor computations, their presence only means that the tensor product is mapped to a repeated sequence of matrix products~\cite{Harrison_2016_ten1}, which does not change the basic approach of sparse-tensor multiplication. Thus, for simplicity only inner/outer products will be considered, explaining why the first and second operands of \Table~\ref{tbl:mult_terms} are single matrices.
\begin{table}
\centering
\footnotesize
  \mycaption{Multiplication Notation}{}
  \begin{tabular}{|C{.3\textwidth}|C{.10\textwidth}||C{.3\textwidth}|C{.10\textwidth}|}
     \hline
       First Operand & $\mathbf{A}$ & Second Operand  & $\mathbf{B}$\\
     \hline
     Rows and Columns of $\mathbf{A}$ & $m$ and $k$ & Rows and Columns of $\mathbf{B}$ & $k$ and $n$ \\
     \hline
     Number of Columns of $\mathbf{A}$ with one or more non-zeros & $\nzc_{\mathbf{A}}$ & Number of Rows of $\mathbf{B}$ with one or more non-zeros & $\nzr_{\mathbf{B}}$ \\
     \hline
     Number of Rows of $\mathbf{A}$ with one or more non-zeros & $\nzr_{\mathbf{A}}$ & Number of Columns of $\mathbf{B}$ with one or more non-zeros & $\nzc_{\mathbf{B}}$ \\
     \hline
   \end{tabular}
  \label{tbl:mult_terms}
\end{table}
Apart from the notation of \Table~\ref{tbl:mult_terms}, this section will use $f$ to refer to the number of floating-point operations in a multiplication, which is the same for all algorithms. $\mathbf{C}$ will denote the matrix product of $\mathbf{A}$ and $\mathbf{B}$. To make the exposition simpler, the subsection will focus mostly on column-by-column versions of the algorithms, \eg{} \ac{CSC}. As such, $f(i)$ will denote the number of floating-point operations to compute the $i$th column of $\mathbf{C}$ and $\nnz_{\mathbf{C}}(i)$ will denote the resulting \ac{NNZ}. LibNT tests for hyper-sparsity by measuring the ratio of \ac{NNZ} to the dimension in question. For example, the row-sparsity of $\mathbf{A}$ can be tested by measuring whether $m/\nnz_{A}>1$.

To begin the discussion, \Sec\ref{sec:mult_datasets} outlines the dataset used for benchmarking. Afterwards, \Sec\ref{sec:sp_times_sp} focuses on LibNT's implementation of the standard sparse multiplication algorithm. This is followed by \Sec\ref{sec:dcsc} and \Sec\ref{sec:csc_no_accum} which describe specialised algorithms to multiply a column-sparse with a row-sparse matrix and a row-sparse with a column-sparse matrix, respectively. Finally, \Sec\ref{sec:outer_product} describes LibNT's algorithm to multiply two index-sparse matrices.

\subsubsection{Dataset}
\label{sec:mult_datasets}

Datasets used for testing can consist of real-world examples or synthetic datasets, which are parameterised and/or randomly generated. While real-world datasets do exist, \eg{} those used in decomposition techniques applied to networks~\cite{Kolda_2008,Papalexakis_2012}, these datasets do not consist of many examples. Thus, to characterise sparse-multiplication algorithms under different conditions, \eg{} \acp{NNZ}, hyper-sparsities, and dimension sizes, this work uses a synthetic dataset. Nevertheless, we return to the differential operators example in \Sec\ref{sec:sparse_results} to provide comparative benchmarks in an application-based context.

We use a third-order tensor generalisation of the R-MAT recursive graph model~\cite{Chakrabarti_2004}, which can control for dimension size, fill factor, and fill pattern. In the original R-MAT model, the recursive \acp{BEP} are specified for each quadrant. To generalise to a third-order R-TENSOR, the \acp{BEP} must be specified for each octant. \Table~\ref{tbl:r_tensor_probs}
\begin{table}
\centering
\footnotesize
  \mycaption{Base edge probabilities used for the R-TENSOR model in the sparse multiplication experiments with their octant specified in parentheses.}{To add variability into experiment runs, the probabilities were adjusted by additive values drawn from a uniform distribution of $[-.1 \,\,.1]$ and renormalised so that they all sum to $1$.}
  \begin{tabular}{|c|c|c|c|c|c|c|c|c|}
     \hline
     Octant:& $(1,1,1)$ & $(1,1,2)$ &$(1,2,1)$ &$(1,2,2)$ &$(2,1,1)$ &$(2,1,2)$ &$(2,2,1)$ &$(2,2,2)$ \\
     \hline
     \acs{BEP}: & $.3$ & $.5/6$ & $.5/6$ & $.5/6$ & $.5/6$ & $.5/6$ & $.5/6$ & $.2$\\
     \hline
   \end{tabular}
  \label{tbl:r_tensor_probs}
\end{table}
outlines the probabilities used for this work. The symbols $a_{ijk}$ and $b_{ijk}$ will be used to denote R-TENSORs. R-TENSORs can manifest column-, row-, or index-sparsity depending on their fill factor and how they are flattened. Maximum \acp{NNZ} were limited by what our workstation could handle when using the \ac{CSC}/\ac{CSR} formats at the highest levels of hyper-sparsity. This allows us to demonstrate the benefits of specialised hyper-sparse formats even when settings allow the use of standard formats. 



\subsubsection{Standard \acs{CSC}/\acs{CSR}}
\label{sec:sp_times_sp}

The tried-and-tested column-by-column \ac{CSC} algorithm relies on a dense, size $k$, singly-compressed array to quickly access columns and a dense, size $m$, accumulator array to quickly collect non-zeros as each column of $\mathbf{C}$ is constructed. Readers unfamiliar with these algorithms and requisite index and accumulator arrays are encouraged to consult Davis~\cite{Davis_2006} and Bulu\c{c} \etal{}~\cite{Buluc_2011b}. \Table~\ref{tbl:characteristics}
 \begin{table}
\centering
\footnotesize
  \mycaption{Characteristics of the three different column-by-column multiplication algorithms. \emph{Importantly, run times include the cost to convert to the \ac{MDT} from column-major \ac{LCO} data}. Memory use only includes temporary data structures used for multiplication.}{}
  \begin{tabular}{C{.11\textwidth}|C{.13\textwidth}C{.12\textwidth}C{.29\textwidth}C{.16\textwidth}}
     \toprule
       Algorithm & Sparse Accumulator & Column Indexing & Run time & Memory Use \\
       \midrule
       \acs{CSC} &yes & singly-compressed & $\mathcal{O}(m+k+\nnz_{\mathbf{A}}+\nnz_{\mathbf{B}}+f+\sum_{i}^{n}\nnz_{\mathbf{C}}(i)\log\nnz_{\mathbf{C}}(i))$ & $\mathcal{O}(m+k)$\\
       \hline
       \acs{DCSC} &yes & doubly-compressed & $\mathcal{O}(m+\nnz_{\mathbf{A}}+\nnz_{\mathbf{B}}+f+\sum_{i}^{n}\nnz_{\mathbf{C}}(i)\log\nnz_{\mathbf{C}}(i))$ & $\mathcal{O}(m+\nzc_{\mathbf{A}})$\\
       \hline
       \acs{CSCNA} &no & singly-compressed & $\mathcal{O}(k+\nnz_{\mathbf{A}}+\nnz_{\mathbf{B}}+\sum_{i}^{n}f(i)\log f(i))$ & $\mathcal{O}(\max f(i)+k)$\\
     \bottomrule
   \end{tabular}
  \label{tbl:characteristics}
\end{table}
summarises the salient characteristics of the \ac{CSC} algorithm.



When \emph{both} matrices present no hyper-sparsity, LibNT opts for the \ac{CSC} or \ac{CSR} algorithms. LibNT gains additional efficiency by converting only one of the matrices to compressed form. For instance, the \ac{CSC} algorithm only requires fast access of the columns of $\mathbf{A}$, meaning it suffices if $\mathbf{B}$ is simply stored in column-major \ac{LCO} format. The primary consideration to choose between the two algorithms is based on what minimises any extra sorts. For instance, if the lexicographic order of both tensors happened to produce column-major matrices once they were flattened, then the \ac{CSC} algorithm will be chosen. In cases were both flattened matrices must be rearranged, the choice is based on a simple heuristic of run time costs of the \ac{CSC} and \ac{CSR} algorithms. Run time between the two is almost identical, except for the final summation term, where the \ac{CSR} algorithm must sort each row of $\mathbf{C}$ instead of each column. Assuming somewhat uniform distribution of non-zeros across rows and columns, the run time for the sort should be smaller if the task is broken into a greater number of pieces. Thus, LibNT opts for the \ac{CSC} format when $n>m$, otherwise it chooses \ac{CSR}.

Finally, by avoiding converting one of the matrices to compressed form, the applicability of the standard algorithms can be extended to greater numbers of cases. For instance, as long as $\mathbf{A}$ has no hyper-sparsity, the standard \ac{CSC} algorithm can be applied, regardless whether $\mathbf{B}$ is row-, column-, or index-sparse. Thus, the \ac{CSC} and \ac{CSR} algorithms can be employed beyond the sparse-sparse case, explaining the first column and row of \Table~\ref{tbl:mult}.

\subsubsection{\acs{DCSC}/\acs{DCSR}}
\label{sec:dcsc}

Hyper-sparsity challenges the \ac{CSC}/\ac{CSR} algorithms in two manners. The first corresponds to cases where using singly-compressed arrays for row or column access are no longer tenable. For instance, should two cubic R-TENSORS be multiplied using
\begin{align}
    a_{ijk}b_{\ell jk} \textrm{,} \label{eqn:rmat_colsparse}
\end{align}
the left and right operands would be mapped to very-wide and very-tall matrices, respectively. Consequently, when $N^{2} \gg \nnz$ using the \ac{CSC} or \ac{CSR} datatypes is prohibitive or even intractable.

A solution is offered by Bulu\c{c} and Gilbert~\cite{Buluc_2008}, who introduced the \ac{DCSC} and \ac{DCSR} formats, which remove run time and memory-use dependance on $k$. Thus, their column-by-column multiplication algorithm can be executed when $\mathbf{A}$ is column-sparse. Similarly, the \ac{DCSR} algorithm can handle cases when $\mathbf{B}$ exhibits row sparsity. Even when $k$ still fits comfortably within memory, the doubly-compressed scheme can produce highly significant speedups. Both variants come at the cost of additional memory accesses, compared to the standard \ac{CSC} and \ac{CSR} options, and so they are not used when the level of hyper-sparsity is likely insufficient to reap the benefits.

To demonstrate these points, experiments used the \ac{CSC} and \ac{DCSC} algorithms to compute \eqref{eqn:rmat_colsparse}. The tests used cubic R-TENSORs generated with \acp{NNZ} ranging from $4e5$ to $1e6$ in increments of $1e5$. Column-sparsity of the R-TENSORs ranged from $1$, \ie{} no column-sparsity, to $500$, \ie{} $1$ non-zero per $500$ columns, in log10-scale increments. The \ac{NNZ} and column-sparsity govern the corresponding dimensions of the R-TENSOR. This was performed $3$ times for each \ac{NNZ}/column-sparsity combination. Finally, two different types of runs were performed. The first run used two different R-TENSORs in \eqref{eqn:rmat_colsparse} and the second used the same R-TENSOR for each operand.

\begin{figure}
  \begin{tabularx}{\textwidth}{YYY}
  \includegraphics{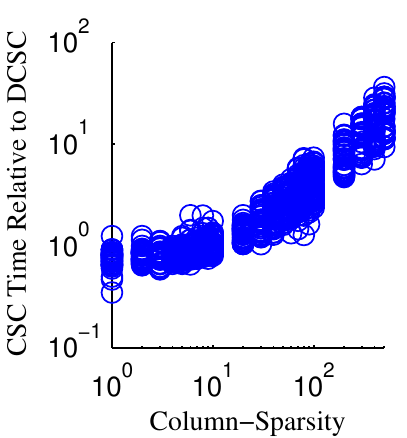} & \includegraphics{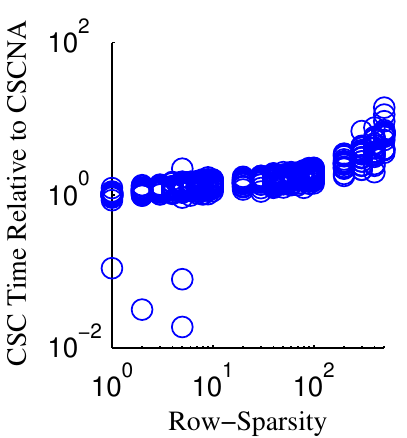} &  \includegraphics{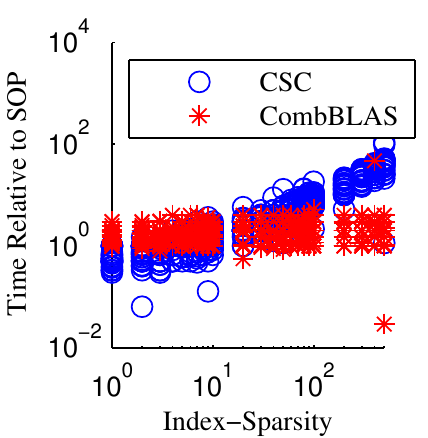}\\
  (a) & (b) &(c)
   \end{tabularx}
  \mycaption{Run times of the multiplication algorithms under differing levels and types of hyper-sparsity.}{(a) graphs the ratio of running-times of the \acs{CSC} vs. the \acs{DCSC} algorithm under different levels of \emph{column-sparsity}, while (b) measures the ratio of the \acs{CSC} vs. the \acs{CSCNA} algorithms under different levels of \emph{row-sparsity}. (c) graphs the ratio of the \ac{CSC} and CombBLAS algorithms to LibNT's \acs{SOP} algorithm under different levels of \emph{index-sparsity}. For all scenarios, performance ratios correlated the most with hyper-sparsity levels.}
  \label{fig:laplace_colsparse}
\end{figure}

Differences in run time were primarily dependent on the hyper-sparsity, and not dimensionality. The relative run times across different levels of hyper-sparsity are depicted in \Fig~\ref{fig:laplace_colsparse}(a). As the figure demonstrates,
a column-sparsity value of $10$ separates the point at which the \ac{DCSC} algorithm outperforms \ac{CSC}. As column-sparsity increases, the \ac{DCSC} algorithm's run time is on average roughly $20$ times faster than the \ac{CSC} approach, demonstrating a tremendous amount of speedup.

For the purposes of LibNT's poly-algorithm, the library opts for the \ac{DCSC} algorithm whenever column sparsity exceeds $3$. While lower than the threshold indicated by \Fig~\ref{fig:laplace_colsparse}(a), this satisfies the primary consideration of keeping memory use proportional to the \acp{NNZ}. LibNT uses the same criteria explained in \Sec\ref{sec:sp_times_sp} to choose between the \ac{DCSC} and \ac{DCSR} variants. As well, and indicated in \Table~\ref{tbl:mult}, LibNT uses the \ac{DCSC} algorithm whenever $\mathbf{A}$ is column-sparse and for all cases of $\mathbf{B}$, except when the latter presents no hyper-sparsity. The reverse holds true for the \ac{DCSR} algorithm.

\subsubsection{\acs{CSCNA}/\acs{CSRNA}}
\label{sec:csc_no_accum}

\Sec\ref{sec:dcsc} outlined a multiplication strategy to handle cases when using the dense \ac{CSC} and \ac{CSR} access arrays becomes untenable. In the opposite scenario, \ie{} multiplying a very-tall matrix with a very-wide one, the dense access arrays of the \ac{CSC} and \ac{CSR} data structures pose no problems and it is the accumulator array that can become untenable. For example, this situation would manifest should two R-TENSORs be multiplied using
\begin{align}
    a_{ijk}b_{\ell m k} \textrm{.} \label{eqn:rmat_rowsparse}
\end{align}

In this situation, the standard \ac{CSC} and \ac{CSR} algorithms can be modified to forego the accumulator array, resulting in the \ac{CSCNA} and \ac{CSRNA} algorithms, respectively. In the column-by-column case, jettisoning the accumulator array means that as each column of $\mathbf{A}$ is constructed the non-zeros are not collected and summed together in one step. Instead for each column $i$, $f(i)$ values are computed and stored in a simple \ac{LCO} list. These $f(i)$ values must then be sorted and any data values with the same \ac{LIV} are then summed together. As \Table~\ref{tbl:characteristics} indicates, this results in an increased sorting burden, but comes at the benefit of not having memory use and run time be dependent on the potentially huge number of rows of $\mathbf{A}$. Note that in this scenario, it is possible to also perform an inner-product algorithm~\cite{Buluc_2011b}. However, due to excessive run times, discussed in more detail in our supplemental material, we do not include its results in our graphs.

As with the \ac{DCSC} algorithm, improvements can be garnered even when the very-large dimensions fit comfortably in memory. To test this, the R-TENSOR experiments in \Sec\ref{sec:dcsc} were repeated, except that \eqref{eqn:rmat_rowsparse} was computed. \acp{NNZ} ranged from $1e5$ to $3e5$ in increments of $5e4$. Apart from this change, all other test settings were kept identical. \Fig~\ref{fig:laplace_colsparse}(b) depicts the results of this test, graphing the ratio of run times of the \ac{CSC} algorithm to the \ac{CSCNA} algorithm under different levels of row-sparsity. As the figure demonstrates, apart from some outliers at low-levels of row-sparsity, the \ac{CSCNA} algorithm is able to match or exceed the \ac{CSC} algorithm. At row-sparsity values of roughly $3$ or higher the \ac{CSCNA} algorithm begins to exhibit faster execution speeds than the \ac{CSC} algorithm, eventually running on average $6$ times faster. Nonetheless, in isolated instances the \ac{CSCNA} algorithm performed considerably worse. Characterising when these situations occur is an important area for further investigation. Even so, as the \ac{CSCNA} algorithm posts excellent performance for the far majority of trials and avoids having memory use and run time depend on $m$, LibNT opts for the \ac{CSCNA} approach whenever row-sparsity is greater than $3$.

As before, LibNT uses the same criteria explained in \Sec\ref{sec:sp_times_sp} to choose between the \ac{CSCNA} and \ac{CSRNA} variants. LibNT also opts for the \ac{CSCNA} algorithm whenever $\mathbf{A}$ is row-sparse and $\mathbf{B}$ is index-sparse while the \ac{CSRNA} is chosen when $\mathbf{B}$ is column-sparse and $\mathbf{A}$ is index-sparse.


\subsubsection{\acs{SOP}}
\label{sec:outer_product}

\begin{figure}
  \centering
  \includegraphics{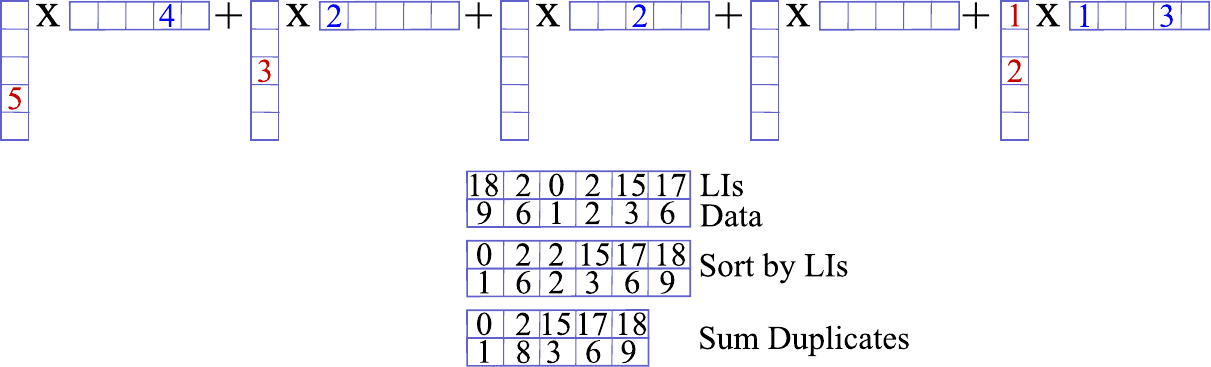}\\
  \caption{The outer-product multiplication algorithm. Using example matrices, the top of the figure demonstrates how the outer-product algorithm would multiply each column of $\mathbf{A}$ with each row of $\mathbf{B}$. Below, the \acs{SOP} algorithm used by LibNT to sum each of the $k$ rank-$1$ matrices is illustrated.}\label{fig:outer_formats}
\end{figure}

The final case to consider is when both $\mathbf{A}$ and $\mathbf{B}$ are index-sparse. Bulu\c{c} and Gilbert~\cite{Buluc_2008,Buluc_2011} have demonstrated that the outer-product approach is fast and memory efficient for this scenario. Under this approach, $\mathbf{A}$ and $\mathbf{B}$ must be sorted in different lexicographical orders---column- and row-major, respectively. As the top of \Fig~\ref{fig:outer_formats} demonstrates,
each column of $\mathbf{A}$ is multiplied with each row of $\mathbf{B}$, producing $k$ $m\times n$ rank-$1$ matrices. To produce the final result, these rank-$1$ matrices must be summed together. Under an index-sparse setting, both $\nzc_{\mathbf{A}}$ and $\nzr_{\mathbf{B}}$ are each less than $k$, and not all non-zero columns of $\mathbf{A}$ have a matching non-zero row of $\mathbf{B}$. Thus, the number of rank-$1$ matrices to sum together is always less than $k$ and often less than $\min(\nzc_{\mathbf{A}}, \nzr_{\mathbf{B}})$.

As part of their CombBLAS library, Bulu\c{c} and Gilbert use their \ac{DCSC} and \ac{DCSR} formats in combination with a heap-like data structure to merge the rank-$1$ matrices~\cite{Buluc_2011}. Bulu\c{c} and Gilbert's motivating problem is large-scale parallel matrix multiplication. As such, the authors correctly did not account for the time to construct doubly-compressed matrices in their performance metrics, because their parallel algorithm amortises these costs across sub-tasks. In contrast, the conversion costs to the \ac{MDT} must be considered in a sparse-tensor setting.

As a result, approaches with minimal conversion costs should also be considered. As the bottom of \Fig~\ref{fig:outer_formats} illustrates, one approach, which we call the \ac{SOP} algorithm, is to just directly concatenate all the intermediate rank-$1$ matrices together into a length $f$ \ac{LCO} data-structure. This \ac{LCO} array can then be sorted based on the \acp{LIV}, with duplicate entries being summed together. The downside is the $\mathcal{O}(f)$ memory use and a $\mathcal{O}(f\log{}f)$ sort, which dominates the complexity. However, unlike in regular sparse settings, $f$ can often be small compared to the \ac{NNZ} of $\mathbf{A}$ and $\mathbf{B}$. Moreover, the ratio of $f$ to the \ac{NNZ} of $\mathbf{C}$ can also be close to $1$, meaning strategies to efficiently add duplicate entries do not always justify their overhead.

These conclusions are borne out when multiplying R-TENSORs using a very similar setup as the experiments in \Sec\ref{sec:dcsc}. However, instead of cubic R-TENSORs, $M\times N \times N$ tensors are used instead, where $M=N^{2}$. Thus, when flattening the R-TENSORs to compute \eqref{eqn:rmat_colsparse}, the resulting matrices exhibit square dimension sizes of $N^{2}\times N^{2}$, providing appropriate conditions to vary the row- and column-sparsity together. Additionally, the \acp{NNZ} varied from $5e4$ to $1e5$ in increments of $1e4$ and the time taken for the \ac{CSC}, \ac{SOP}, and CombBLAS'~\cite{CombBLAS} C++ index-sparse algorithm, including setup costs, was measured. All other conditions were kept the same.

The results of this test are depicted in \Fig~\ref{fig:laplace_colsparse}(c). The \ac{SOP} algorithm outperformed the \ac{CSC} algorithm at most of the  index-sparsity range. Inspection of the numerical results reveal that faster run times begin at index-sparsities greater than $6$, with the performance gap increasing to roughly $35$ times faster execution at the highest levels of index-sparsity. Compared to the CombBLAS algorithm, the \ac{SOP} outperformed it on average by a factor of $2$ at all levels of index-sparsity, demonstrating the value of a simplified approach in sparse tensor settings. However, in several instances CombBLAS outperformed the \ac{SOP} algorithm, indicating that certain scenarios call for a more sophisticated merging approach. Further work should focus on identifying these scenarios \emph{a priori}. Even so, the \ac{SOP} executed the fastest on almost all test instances.

As a result, LibNT opts for the \ac{SOP} algorithm whenever both $\mathbf{A}$ and $\mathbf{B}$ are index-sparse. For the purposes of satisfying the primary consideration of avoiding highly excessive memory use, the \ac{SOP} algorithm is applied whenever both row- and column-sparsity exceed $3$.

%
%
%
%

\section{Comparative Performance}
\label{sec:sparse_results}

So far, this work has contrasted performance of different algorithmic choices. What has not been discussed is the impact of using such high-performance kernels in an actual multi-purpose sparse-tensor arithmetic setting. Returning to the differential operators example of \Sec\ref{sec:prelim} and using left-to-right precedence, the first term in \eqref{eqn:isotropic_new} can be broken into two binary products, with the second written as
\begin{align}
  b^{(y)}_{ii'j\ell}d^{(y)}_{i'k} \textrm{,}  \label{eqn:hyper_sparse_app}
\end{align}
where $b^{(y)}_{ii'j\ell}=d^{(y)}_{ii'}\delta_{j\ell}$. Assuming uniform index ranges, when flattened \eqref{eqn:hyper_sparse_app} describes an $N^{3}\times N$ row-sparse matrix as the left operand multiplied with a standard sparse $N\times N$ matrix.

To unearth some of the significance of using this work's techniques, we compare the performance of NTToolbox against two alternatives in executing \eqref{eqn:hyper_sparse_app}. The first alternative is the \ac{MTT}. Like the NTToolbox, the \ac{MTT} relies on a MATLAB frontend to setup and call optimised compiled-language backends, except that the former uses LibNT's C++ algorithms specialised for tensor arithmetic, while the latter relies on MATLAB's optimised, but general-purpose, built-in routines. Differences in performance will be partly driven by any limitations that the MATLAB environment imposes upon the \ac{MTT}. The \ac{MTT} handles all hyper-sparsity combinations by executing \ac{CSC} multiplication after excision of all-zero rows and columns. As such, the number of sorts performed remains constant, \ie{} through \ac{MTT}'s three and two calls to MATLAB's \texttt{unique} function and \ac{CSC} sparse matrix construction routines, respectively. A head-on comparison helps uncover when NTToolbox's specialised high-performance kernels are warranted over the convenience of a pure MATLAB implementation.

The second alternative is an NTToolbox re-implementation of the \ac{MTT}'s multiplication strategy, using a LibNT-based backend. For fair comparison, care was taken in optimising this approach, \eg{} only converting the right operand to the \ac{CSC} \ac{MDT} and avoiding the superfluous excision of all-zero columns of the right operand. Thus, differences in performance will be driven solely by the impact of the poly-algorithm vs. the all-purpose strategy of \ac{CSC} multiplication combined with excision used in the \ac{MTT}.

\begin{figure}
  \centering
  \includegraphics[scale=.92]{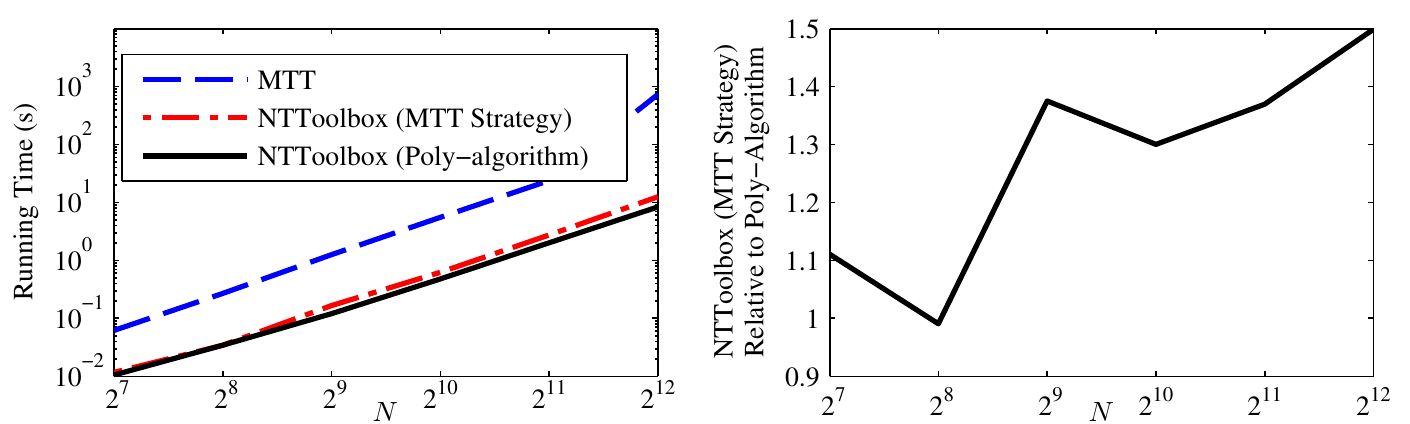}
  \mycaption{Comparative performance of NTToolbox multiplication poly-algorithm vs. the \ac{MTT} and an NTToolbox implementation of the \ac{MTT}'s multiplication strategy.}{All graphs depict the run time to execute the hyper-sparse products in \eqref{eqn:isotropic_new}. (a) graphs all three run times in logarithmic scale, whereas (b) graphs the run time of the NTToolbox (MTT strategy) vs. the NTToolbox (poly-algorithm).}
  \label{fig:mult_compare_differential}
\end{figure}

Tests assumed input images of size $N \times N$, where $N=2^{k}-1$, with increasing values of $k$. The subtraction of $1$ ensures the integer division libraries used by NTToolbox are not given an unfair advantage.  The time needed to rearrange both operands into the required lexicographical order for the poly-algorithm was included in the measurements. \Fig~\ref{fig:mult_compare_differential}(a) graphs the run times of all three implementations,
demonstrating that NTToolbox speeds up the calculation by an order of magnitude or more. \Fig~\ref{fig:mult_compare_differential}(b) depicts the run time of the NTToolbox (MTT strategy) relative to the NTToolbox (poly-algorithm), demonstrating that the latter speeds up computations by $40\%$ or more at high dimensionalities. For context when $N=2^{11}-1$ the run time of the \ac{MTT}, NTToolbox (MTT strategy), and NTToolbox (poly-algorithm) were $23.7\,s$, $2.7\,s$, and $2.0\,s$, respectively. For the NTToolbox-specific tests, the performance improvements of the poly-algorithm can mainly be attributed to eliminating superfluous permutations, meaning only those required to flatten operands into matrix form are executed.

\section{Conclusion}
\label{sec:sparse_conclusion}

LibNT offers a multi-purpose environment for the sparse tensor arithmetic operations seen in Einstein-like notation, meaning addition, subtraction and multiplication, and the solution of equations incorporating dense, sparse, or dense/spase tensor mixtures. This work focused on three core aspects. First, like Bader and Kolda~\cite{Bader_2007}, we believe a multi-purpose arithmetic library should not place an \emph{a priori} precedence on certain tensor indices over others. However, we argue for the \ac{LCO} format over Bader and Kolda's \ac{CO} format, presenting results showing faster sort run times. Importantly, these benefits come with a smaller memory footprint, especially at higher orders. Currently tensor dimensionalities are limited to $63$ bits, but future work incorporating very-large integer datatypes should remove this limitation, extending the \ac{LCO}'s benefits to a greater set of tensor problems.

Secondly, we emphasise the importance of high-performance rearrangement algorithms when using list-like data structures such as the \ac{CO} and \ac{LCO} formats. Such algorithms are necessary to realise a high-performance sparse tensor arithmetic library. This work outlined the impact of using radix sort, which is specialised to sort integer datatypes, over more general-purpose sorting algorithms. Importantly, we also outline how to take advantage of the inherent structure of sparse data to speed up the frequent permutations required for list-like data structures. An algorithm exploiting these underlying characteristics was developed, outperforming the fastest standard sorting option and demonstrating the value of employing specialised approaches to sparse tensor arithmetic.

Finally, we addressed how to implement sparse-times-sparse tensor multiplication, an operation that exemplifies the unique requirements of sparse tensor arithmetic. A multi-purpose library could encounter any combination of sparse, index-sparse, column-sparse, or row-sparse data, which all demand their own specialised approaches. Apart from highlighting this unique characteristic, we also outlined a multiplication poly-algorithm that can choose appropriate algorithms accordingly. The poly-algorithm ensures that excessive memory use is avoided, a potentially catastrophic event. Moreover, the poly-algorithm produces highly-significant reductions in run time over the common \ac{CSC}/\ac{CSR} approach. While the \ac{MTT} can also handle hyper-sparsity, its one-size-fits-all algorithm does not exploit the very distinct features of the different sparsity types.

We demonstrate the impact of this work on several benchmarks derived from the application of high-order differential operators. These tests are complemented by other benchmarks incorporating randomly generated sparse tensors. Compared to the \ac{MTT}, the outlined kernels contributed to considerable improvements in run time on constructing high-order combinatorial Laplacians, demonstrating the value of this work's specialised and high-performance kernels. The discussed high-performance kernels are accessible through the LibNT and NTToolbox, which are open-source libraries for Einstein-like notation, implemented in C++ and MATLAB, respectively. However, the data structures and algorithms described here, or variants thereof, are also well suited to any other package incorporating sparse tensor arithmetic.

Considerable future work can further advance the state of sparse-tensor arithmetic. In particular, multi-core and heterogeneous routines would be welcome, \eg{} parallel approaches to radix sort~\cite{Satish_2010}. Another possibility is to leverage work within the graph algorithm community on parallel sparse matrix-matrix multiplication~\cite{Buluc_2011,Buluc_2012}. Efficiencies stemming from symmetry of sparse tensors, possibly adapting existing dense approaches~\cite{Schatz_2014}, should also be incorporated. Reducing temporary memory allocation in chained arithmetic expressions should help reduce overhead. Finally, adapting LibNT to be able to handle very-large integers as \acp{LIV} is highly important. These and other advancements will help further the impact of this, and other~\cite{Bader_2007,Parkhill_2010}, efforts towards establishing a mature body of sparse tensor arithmetic routines.

\bibliography{docutex-minimal}
\bibliographystyle{siam}

\section{Supplemental Material}

\subsection{Sorting}
\label{sec:sorting_sparse}

We provide more details on the performance of \ac{MSD} radix sort vs competitor algorithms in sorting sparse tensor data. When sorting \acp{LIV}, another array, \ie{} the data array, must be sorted alongside it. By increasing the memory bandwidth needed to perform rearrangements, cache misses can be more predominant, which is often the leading factor in sorting performance~\cite{LaMarca_1999}. The \ac{LCO} sorting task can be categorised as sorting with ``satellite'' data~\cite{Cormen_2001}, which is often tackled using pointers to the additional data or by using data structures containing key/record pairs. In contrast, sorting \ac{LCO} data requires sorting a contiguous integer-valued \ac{LIV} array with a contiguous and separate data array acting as satellite data. The importance of fast sorting speed to the sparse \ac{LCO} format makes it highly worthwhile to focus on optimising sorts.

We tested and adapted several different algorithms using our own C++ implementations. The leading comparison-based sorting algorithms that were tested include the highly prominent introspective sort~\cite{Musser_1997}, adapted from Schwarz~\cite{Schwarz_2006}, and Timsort~\cite{timsort}. Integer sorting algorithms were also tested, including a \ac{LSD} radix sort~\cite{Sedgewick_1998} and an in-place version of \ac{MSD} radix sort~\cite{MSD_radix_inplace} that uses no extra memory. All of the tested algorithms were adapted to sort the \ac{LCO} data array alongside any movements of the \ac{LCO} \ac{LIV} array. In addition, effort was taken to optimise their implementations, including using hybrid and adaptive approaches, in order to achieve fast run times.

%

\begin{figure}
  \center
  \begin{tabular}{cc}
     \includegraphics{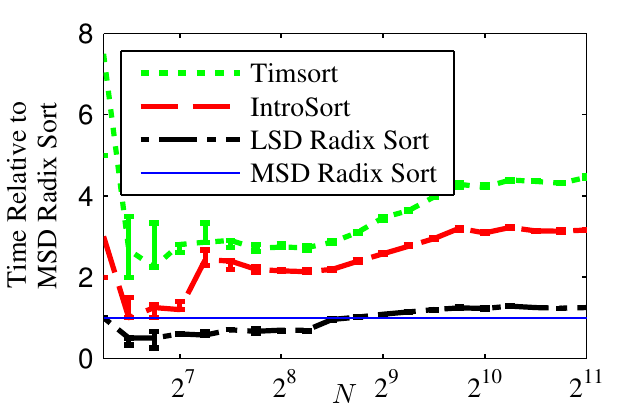} & \includegraphics{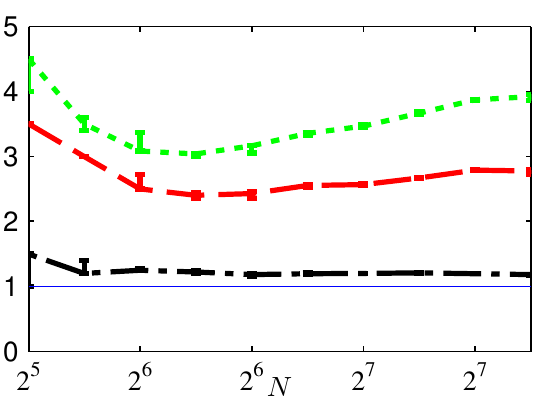} \\
     (a) & (b)
  \end{tabular}
  \mycaption{Benchmark results of different sorting algorithms applied to sparse fourth-order tensors.}{(a) and (b) depict results from randomly generated $N\times N \times N \times N$ sparse tensors, with a fill factor of a fourth-order Laplacian and a $5\%$ fill factor, respectively. Tests were run on increasing values of $N$ and repeated $10$ times. Trend lines represent median values and error bars represent quartiles.}
  \label{fig:sorting_times}
\end{figure}

Two different types of tests were performed. The first type of test, depicted in \Fig~\ref{fig:sorting_times}(a),
measures the sorting time on a randomly-generated tensor matching the fill-factor of a fourth-order Laplacian operator, \eg{} one that can act on an image. The Laplacian operator's fill factor decreases quadratically with dimensionality, providing a highly-sparse test setting. While it is important to measure performance in a highly-sparse setting, it is also worthwhile to test under settings where sparsity does not vary quadratically with dimensionality. Along those lines, \Fig~\ref{fig:sorting_times}(b) depicts sorting times of fourth-order tensors with $5\%$ fill factors.

As the figure makes clear, both radix sorts beat out the two comparison sorts in a highly-sparse setting, posting $2$ to $4$ times faster speeds for most of the range of dimensionalities. These results are more striking when considering that the benchmark setup is directly unfavourable to radix sorts. More specifically, a fourth-order Laplacian's fill factor decreases quadratically, meaning the maximum size of the \ac{LIV} increases at a quadratic rate compared to the linear rate increase of the \ac{NNZ}. This can be problematic for radix sorts, because its run time is proportional to the magnitude of the keys being sorted~\cite{Sedgewick_1998}, \ie{} the \acp{LIV}. Nonetheless, these results indicate that radix sort can perform extremely well even in this demanding setting. One likely reason for this is that both radix sort variants used a commonly recommended~\cite{Sedgewick_1998} hybrid implementation that switched to a comparison-based sort when appropriate. Thus, dependence on the magnitude of the \acp{LIV} is relaxed.

Similar results were produced when the algorithms were tested on sparse tensors with $5\%$ fill factor, with the radix sorts outperforming their comparison counterparts by highly significant margins.


Apart from illustrating the high-performance of radix sorts, these results demonstrate the significant impact of algorithm choice in sorting sparse tensors. Depending on the choice of algorithm, sorting can take roughly $2$-$4$ times longer, which is of high consequence when considering the importance of rearrangement operations to sparse tensor computations. In terms of whether the \ac{LSD} or \ac{MSD} variant is preferable, the latter generally outperformed the former, particularly at very-large values of $N$. Moreover, the \ac{MSD} version used here is inplace. For these reasons, LibNT uses \ac{MSD} radix sort.

\subsection{Excising All-zero Rows and Columns}

We provide more details on the rationale for why we avoid excising all-zero rows and columns and use instead specialised algorithms designed to handle hyper-sparsity. To help make our explanation as concrete as possible, we use an example where two tensors have been flattened into matrices, $\mathbf{A}$ and $\mathbf{B}$, to execute a tensor product. These are multiplied together in the equation $\mathbf{A}*\mathbf{B}$. First, we will assume $\mathbf{A}$ is row-sparse and proceed through the four possible hyper-sparsity options of $\mathbf{B}$. Second we will assume $\mathbf{B}$ is row-sparse and proceed through the four possible hyper-sparsity options of $\mathbf{A}$

Before beginning with the first case, \ie{} assuming $\mathbf{A}$ is row-sparse, we note that one approach to handle $\mathbf{A}$'s row-sparsity, regardless of the hyper-sparsity of $\mathbf{B}$, is to sort $\mathbf{A}$ in row-major order, excise the all-zero rows, and then re-sort $\mathbf{A}$ back in column-major order. $\mathbf{B}$ would be sorted once into column-major order. In this scenario, \ac{CSC} multiplication can be executed, which can be done by only converting the excised version of $\mathbf{A}$ into the \ac{CSC} format and leaving $\mathbf{B}$ in \ac{LCO} format, which avoids any possible issues should $\mathbf{B}$ be hyper-sparse in any way. This approach, however, requires an extra sort, which is why we avoid this option.

Going through the four possible hyper-sparsity characteristics of $\mathbf{B}$ brings up the following considerations:

\begin{enumerate}
  \item \emph{$\mathbf{B}$ is simply sparse}. In this case, we can sort both matrices in row-major order, excise all-zero rows of $\mathbf{A}$, and then perform standard \ac{CSR} multiplication. However, in \ac{CSR} multiplication, only $\mathbf{B}$ need be in compressed form. So in this scenario, only $\mathbf{B}$ need be converted to \ac{CSR} form, and $\mathbf{A}$ can be kept in row-major \ac{LCO} form. This approach requires no excisions.
  \item \emph{$\mathbf{B}$ is row-sparse}. The number of columns of $\mathbf{A}$ must match the number of rows of $\mathbf{B}$ to be a valid matrix multiplication. As a result, even though $\mathbf{B}$ is row-sparse, because $\mathbf{A}$ is \emph{not} column-sparse, we know that we can safely store the \ac{CSR} version of $\mathbf{B}$. Thus, there is no need perform any excision, and the standard \ac{CSR} algorithm can be employed, keeping $\mathbf{A}$ in row-major LCO form.
  \item \emph{$\mathbf{B}$ is column-sparse}. There are several options in this case:
    \begin{enumerate}
        \item To stay with the excision approach we can sort both matrices in row-major order and then perform standard \ac{CSR} multiplication. However, because $\mathbf{B}$ is column-sparse, the sparse accumulator used in the \ac{CSR} algorithm can consume amounts of memory far exceeding the \ac{NNZ} of either $\mathbf{A}$ or $\mathbf{B}$. Thus, to perform the \ac{CSR} algorithm, $\mathbf{B}$ would need to be first sorted in column-major order, have its all-zero columns excised, and then be re-sorted in row-major order.  Similar issues apply if we attempt to perform \ac{CSC} multiplication. Thus, an additional expensive sort would be required.
        \item Another option is to sort $\mathbf{A}$ and $\mathbf{B}$ in row- and column-major orders respectively and perform an inner product algorithm~\cite{Buluc_2011b}, which does not require excising all-zero rows and columns. A sparse accumulator can be used to store one-by-one either the non-zero rows of $\mathbf{A}$ or the non-zero columns of $\mathbf{B}$. However, in this case, the run time cost would be either $\mathcal{O}(\nzr_{\mathbf{A}}\nnz_{\mathbf{B}})$ or $\mathcal{O}(\nzc_{\mathbf{B}}\nnz_{\mathbf{A}})$, which is typically on the same order of magnitude as $\mathcal{O}(\nnz_{\mathbf{A}}\nnz_{\mathbf{B}})$. In our experiments, this option ran one to three orders of magnitude slower than the \ac{CSCNA} or \ac{CSRNA} algorithms. For this reason, we do not include its results in this work.
                \item Due to the above considerations, we use a variant of the \ac{CSC} and \ac{CSR} algorithms, \ac{CSCNA} and \ac{CSRNA}, respectively, that eschew the sparse accumulator, thus avoiding the need for additional re-sorts or the use of the expensive inner-product algorithm. The choice of \ac{CSCNA} vs. \ac{CSRNA} is based on the criteria given in Section 4.2.3. Either way no excisions are required.
    \end{enumerate}
  \item \emph{$\mathbf{B}$ is index-sparse}. Similar issues as the previous case ensue, except with even more complications. However, since $\mathbf{A}$ is only row-sparse, we can use the CSCNA algorithm and data structure.
\end{enumerate}

To help complete this picture, we also outline the issues that arise if $\mathbf{B}$ is row-sparse and proceed through the following four hyper-sparsity options of $\mathbf{A}$:

\begin{enumerate}
 \item \emph{$\mathbf{A}$ is simply sparse}. This is similar to 1 in the first set of considerations, except that we perform \ac{CSC} multiplication instead.
 \item \emph{$\mathbf{A}$ is column-sparse}. For simplicity, we outline the options involving algorithms relying on row-major order, but identical considerations apply in the column-major case. It is possible to sort both matrices in row-major order, excise the all-zero rows of $\mathbf{B}$, and then perform CSR multiplication. However, in this case, excising the all-zero rows of $\mathbf{B}$ is a challenge, as it requires coordinating in some way with the columns of $\mathbf{A}$, since the columns and rows of $\mathbf{A}$ and $\mathbf{B}$, respectively, are inner-product dimensions and any matchings before pre-excision must remain post-excision. For this reason the rows of $\mathbf{B}$ cannot be excised independently. So to stay with the excision approach there are two options.
      \begin{enumerate}
      \item One option would be to sort $\mathbf{A}$ in column-major order to excise the columns matching the excised rows of $\mathbf{B}$. $\mathbf{A}$ would then be re-sorted in row-major form and \ac{CSR} can be executed. This would require an additional expensive sort.
      \item Another option is to avoid excising the matching columns of $\mathbf{A}$ and instead use some form of on-the-fly mapping of the columns of $\mathbf{A}$ so that they match the excised versions of the rows of $\mathbf{B}$. This would require creating an additional data structure, such as a binary search tree or a hash map.
      \item  For either option, when columns of $\mathbf{A}$ are excised there is no guarantee they will be all-zero, which will leave orphaned data and indices. So, temporary memory will be needed of the truncated data and indices of $\mathbf{A}$ (since we want to avoid altering the original tensor data), or the multiplication algorithm would need to be able to handle ``skip'' columns of $\mathbf{A}$.
      \item  All this to say, that excising the all-zero rows of $\mathbf{B}$ requires additional sorts or additional data structures with their own costs. Our approach avoids these complications altogether by using the \ac{DCSR} or \ac{DCSC} algorithms, which do not need to perform any excisions.
      \end{enumerate}
  \item \emph{$\mathbf{A}$ is row-sparse}. Similar considerations as 2 in the first set of considerations apply, except that the \ac{CSC} algorithm should be performed.
  \item  \emph{$\mathbf{A}$ is index-sparse}. Similar considerations as 4 in the first set of considerations apply, except that the \ac{CSRNA} algorithm should be performed.
\end{enumerate}

These considerations generalise to additional hyper-sparsity combinations of $\mathbf{A}$ and $\mathbf{B}$, except for certain cases where the roles of $\mathbf{A}$ and $\mathbf{B}$ are reversed, which would then require using the counterpart of the algorithms described above, \eg{} \ac{CSC} instead of \ac{CSR}.

\end{document}